\documentclass[fleqn,usenatbib,useAMS]{mnras}
 \usepackage{epsfig}
 \usepackage{relsize}
 \usepackage{josty}
\usepackage{amsmath}
\usepackage{hyperref}

\title[Stellar Populations With Optical Spectra] 
{Stellar Populations With Optical Spectra: Deep Learning vs. Popular Spectrum Fitting Codes}
\author[Woo et al.] 
    {\parbox{\textwidth}{Joanna Woo,$^{1}$\thanks{\texttt{j\_woo@sfu.ca}}
    Dan Walters,$^{2}$
    Finn Archinuk,$^{2}$
    S. M. Faber,$^{3}$
    Sara L. Ellison,$^{2}$
    Hossen Teimoorinia,$^{2,4}$
    Kartheik Iyer$^{5}$
    } 
\vspace{0.4cm}\\
\parbox{\textwidth}{ 
$^{1}$Department of Physics, Simon Fraser University, 8888 University Drive, Burnaby BC V5A 1S6, Canada\\
$^{2}$Department of Physics \& Astronomy, University of Victoria, PO Box 1700 STN CSC, Victoria BC V8W 2Y2, Canada\\
$^{3}$University of California Observatories/Lick Observatory,
  Department of Astronomy and Astrophysics, University of California, 
  Santa Cruz, CA 95064, USA\\
$^{4}$NRC Herzberg Astronomy and Astrophysics, 5071 West Saanich Road, Victoria, BC, V9E 2E7, Canada\\
$^{5}$Columbia Astrophysics Laboratory, Columbia University, 550 West 120th Street, New York, NY 10027, USA
}}

\pagerange{\pageref{firstpage}--\pageref{lastpage}} \pubyear{2024}

\begin{document}
\label{firstpage}
\maketitle

\begin{abstract}
We compare the performance of several popular spectrum fitting codes (Firefly, {\sc starlight}, pyPipe3D and pPXF), and a deep-learning convolutional neural network (StarNet), in recovering known stellar population properties (mean stellar age, stellar metallicity, stellar mass-to-light ratio $\MLr$ and the internal E(B-V)) of simulated galaxy spectra in optical wavelengths.  Our mock spectra are constructed from star-formation histories from the IllustrisTNG100-1 simulation.  These spectra mimic the Sloan Digital Sky Survey (SDSS) through a novel method of including the noise, sky residuals and emission lines taken directly from SDSS. 
We find that StarNet vastly outperforms all conventional codes in both speed and recovery of stellar population properties (error scatter $<0.08$ dex, average biases $<0.02$ dex for all tested quantities), but it requires an appropriate training set.  Of the non-machine-learning codes, pPXF was a factor of 3-4 times faster than the other codes, and was the best in recovering stellar population properties (error scatter of $<0.11$ dex, average biases $<0.08$ dex).  However, the errors and biases are strongly dependent on both true and predicted values of stellar age and metallicity, and signal-to-noise ratio.  The biases of all codes can approach 0.15 dex in stellar ages, metallicities and log $\MLr$, but remain $\ltsima 0.05$ for E(B-V).  Using unrealistic Gaussian noise in the construction of mock spectra will underestimate the errors in the metallicities by a factor of two or more, and mocks without emission lines will underestimate the errors in stellar age and $\MLr$ by a factor of two.  

\end{abstract}


\section{Introduction}
\label{introduction}

Determining stellar population properties (such as the mean stellar age and metallicity) of a galaxy based on its integrated optical spectrum has been a challenge for several decades.  Pioneers of methods to derive stellar population properties from spectra (or broad-band spectral energy distributions) include \cite{Morgan1956,Wood1966,Tinsley1968,Faber1972,Spinrad1972}.  See \cite{Conroy2013} for a contemporary overview of the methods, applicability, and the degeneracies inherent in these techniques, especially at optical wavelengths.  The basic premise of ``full spectrum fitting" (or more generally ``population synthesis") is that the integrated spectrum of a galaxy is the linear combination of light from many simple stellar populations (SSPs) of known ages, metallicities and chemical enrichments.  Decomposing a galaxy spectrum into its constituent populations in order to derive the star formation history (SFH), or the average properties of the SFH, is the goal of full spectrum fitting.  

Several modern spectrum fitting codes have been designed to perform this type of decomposition, and many are in widespread use.  {\sc starlight} \citep{CidFernandas2005}, Penalized PiXel Fitting (pPXF - \citealp{Cappellari2004,Cappellari2017}, Firefly \citep{Wilkinson2017} and pyPipe3D \citep{Sanchez2016,Lacerda2022} are four popular examples of modern spectrum fitting codes.

It has been known for some time that optical spectra are degenerate in age, metallicity and dust properties.  For example, the spectra of populations with differing ages can have very different UV components, but look virtually indistinguishable in the optical \citep{LopezFernandez2016}.  Inferences of the posterior distributions of age and metallicity for single galaxies span the entire anti-correlated locus that marks the degeneracy in age and metallicity \citep{Wilkinson2017}.

A promising avenue for helping to resolve such degeneracies is the use of Deep Learning (DL) which is becoming a widespread tool in astronomy.  DL has been used for classifying galaxy images (e.g., \citealp{Huertas-Company2018,Bottrell2019,Bickley2021}), classifying galaxy spectra (e.g., \citealp{Teimoorinia2022}), determining photometric redshifts (e.g., \citealp{Brescia2021}), among many other applications (e.g., \citealp{Bluck2022}).  DL methods use many more parameters than conventional methods, finding its own fitting model that translates fluxes in each pixel of a spectrum to a quantity of interest, such as the mean stellar age or metallicity.  \cite{Lovell2019} applied DL methods to determine star formation histories from optical galaxy spectra with some success.  Although they only tested spectra with rather high S/N (of 50), their study nevertheless motivates a comparison with conventional spectrum fitting methods.  

A troubling sign that all is not well with full spectrum fitting is that different algorithms produce conflicting results for basic scientific questions.  For example, \cite{Zheng2017} running {\sc starlight} on the MaNGA IFU survey find slightly negative age gradients (older centres) with very little correlation with mass.  In contrast, \cite{Breda2020} running {\sc starlight} on CALIFA galaxies, and \cite{Li2018} running pPXF on MaNGA galaxies find gradients that imply older centres for massive galaxies and younger centres for low-mass galaxies.  Correlations with galaxy morphology are also inconsistent.  \cite{Goddard2017} using Firefly on MaNGA find slightly younger centres in early-type galaxies and older centres for late-type galaxies, while \cite{Woo2019} running pPXF on MaNGA find the opposite trend: older centres in quiescent early-type galaxies and a mix of age gradients in late-type galaxies.  \cite{GonzalezDelgado2014} running {\sc starlight} on the CALIFA sample find older centres for most galaxies, but flatter age gradients in early-type galaxies.  The study by \citep{AvilaReese2023} on the evolution of galaxy mass build-up over time is even more ambitious than the others.  Instead of using mere averages of the star formation history, they draw conclusions based on the full detailed histories on the resolution of a 1-3 Gyr. All of these authors made scientific inferences about the evolution of galaxies based on their results.  

The study by \cite{Lu2023} investigated age and metallicity gradients using a single code (pPXF) but with three different sets of model SSPs templates.  Encouragingly, they showed that the broad trends between age and metallicity gradients with velocity dispersion were qualitatively similar between the three sets of templates, suggesting that it is the fitting algorithms rather than the choice of templates that may be responsible for the discrepancies between the above studies.  {However, an older study by \cite{CidFernandes2014} found metallicities can vary widely for different template choices for a single code ({\sc starlight}), but ages, extinctions and masses are relatively robust to the choice of templates.}

What makes it difficult to understand the sources of these discrepancies is the fact that not only are the codes using different fitting algorithms, they also choose different sets of SSP templates to fit, which are constructed from different stellar libraries and sometimes different initial mass functions (IMFs).  An apples-to-apples comparison of the codes is relatively lacking in the community, where each code is tested with identical conditions on an identical set of optical spectra for which the stellar population properties are known.  {One study \citep{Magris2015} compared the recovery of stellar population parameters for four codes: {\sc starlight}, DynBas, TGASPEX and GASPEX, the last three of which are not public.
They found that all codes fit their test spectra with ``almost perfect fits" but that their recovered parameters differed significantly.  
}
Another study by \citep{Ge2018} compared the performance of two codes, pPXF and {\sc starlight}, under the same conditions, finding that pPXF recovers stellar age, metallicity, mass-to-light ratios and dust extinction with smaller errors and biases than {\sc starlight}.  However \cite{CidFernandes2018} claims that the study by \cite{Ge2018} was flawed by non-ideal choices in the range of the colour-excess E(B-V) parameter.  
\cite{Pacifici2023} compare 14 codes that fit broad-band SEDs (not spectra) and find consistencies in stellar mass, but not star-formation rate and dust attenuation.  

The goal of our study is to systematically compare the ability of four popular spectrum fitting codes (Firefly, pPXF, pyPipe3d, and {\sc starlight}) and one DL method (StarNet - \citealp{Fabbro2018}) to recover known stellar population properties from simulated optical spectra.  We specifically test the codes' ability to recover four properties which are commonly measured in the literature: the mean mass-weighted stellar age, the mean mass-weighted stellar metallicity, the $r$-band stellar mass-to-light ratio, and the colour excess E(B-V).

Most codes when they are presented, are published with mock spectra to validate their methods.  These spectra were either created from models or from degraded spectra of globular clusters.  All codes claim excellent performance, often with less than 0.1 dex error in the recovery of stellar population properties such as the mean stellar age and the mean stellar metallicity.  Yet, they are not always in agreement when applied to observational data, as evidenced by the examples given above.  The mock spectra are constructed differently by the authors of the codes, applying different assumptions, noise profiles, and treatment of (or lack of) emission lines.  

What is needed is a standardized set of mock spectra that truly mimic realistic spectra from a target survey (such as the Sloan Digital Sky Survey - SDSS \citealp{York2000a}).  A key novel aspect of our analysis is the use of realistic noise.  What has generally not been appreciated in past studies is that realistic noise is often not Gaussian, with sky residuals being a major contributor to non-Gaussian noise.  In this study, we construct a suite of mock spectra that mimic the SDSS in nearly every way, not only in the distributions of redshifts, signal-to-noise ratios (S/N) and velocity dispersions, but also in the sky residuals, emission lines and even foreground reddening, while the underlying SFHs are from the IllustrisTNG simulation (\citealp{Nelson2019}).  We constructed this suite of spectra with DL in mind: if DL is to eventually be applied to real observed spectra, it must be trained on spectra that actually look like real spectra, although the applicability to real spectra will be explored more deeply in future work.  In principle, our method can be applied to create mock spectra for any data set, such as MaNGA, MUSE etc.  We have chosen to mimic the SDSS simply for its unparalleled volume and richness, public availability, prevalence and longevity in the galaxy evolution community and beyond.

There are many spectrum fitting codes out there that could be compared\footnote{The site sedfitting.org has a long list of them.}, and due to computational and human limitations, we could not test them all.  Our choice of Firefly, pPXF, pyPipe3D and {\sc starlight} was motivated by popularity, computation time and ease of use.  Notable codes that we chose not to test include Prospector \citep{Johnson2021}, FADO \citep{Gomes2017} and BAGPIPES \citep{Carnall2018}.  Prospector is popular but performs a computationally expensive Bayesian exploration of the full parameter space, taking 1-2 days to analyze a single spectrum (S. Tacchella, private communication).  Thus running Prospector is unfeasible for large datasets such as the SDSS.  FADO is one of the few codes that uses emission lines to constrain the SFH.  However, as detailed in \secref{simulatedspectra}, our mock spectra are currently constructed such that emission lines are unrelated to the SFH (due to DL considerations), and are therefore not suitable for testing FADO.   BAGPIPES uses Bayesian analysis to infer the parameters of parametric SFHs.  Since the other codes we tested fit non-parametric SFHs, the accuracy of the stellar population parameter recovery could not be fairly compared with BAGPIPES.  (For a high-$z$ comparison between BAGPIPES and Prospector see \citealp{Kaushal2023}.)

\section{Constructing A Suite of Mock Spectra}
\label{simulatedspectra}

In order to evaluate an algorithm's ability to derive stellar population properties from spectra (namely the stellar age, stellar metallicity, stellar mass-to-light ratio and colour excess), we require a suite of mock spectra for which 1) these properties are known, and 2) which closely mimic the noise properties of observed spectra.  To satisfy the first of these requirements, we extract star-formation histories from 26201
galaxies from between $z=0$ and $z=0.03$ in the IllustrisTNG100-1 simulation (\citealp{Nelson2019} - hereafter TNG), selecting as follows.  First we selected all centrals with $M_* > 10^{9.5} \Msun$ at $z=0$ (12200).
We divided the galaxies into star-forming (SF, 7589),
green valley (GV, 1743),
and quiescent (Q, 2868)
following the same star-formation rate criteria as \cite{Walters2021}. We then randomly discarded 5589 of the star-forming galaxies to roughly balance the types of star-formation histories and metallicity evolution (which we collectively refer to as ``SFH'') in our sample. Although having an even distribution of star-formation histories has no effect on the conventional spectrum fitting codes (since the codes fit each spectrum individually), it has an impact on the training of StarNet, which is a convolutional neural network. If the training sample lacks variety, the network can overfit to the most common examples. Next, we split our $z=0$ sample into a training set ($\sim$80\%) and a test set ($\sim$20\%). To expand our sample, we further selected the galaxies at $z=$ 0.01, 0.02 and 0.03 belonging to the main progenitor branch of our $z=0$ galaxies, assigning each galaxy to a set (training/test) based on their $z=0$ descendant.  It was necessary to follow the main progenitor branch in order to ensure that there was no cross-contamination of galaxies or their progenitors between the training and test samples.  By expanding our sample to include progenitors at earlier adjacent snapshots, we necessarily include very similar, almost duplicate SFHs in both samples.  They would naturally produce nearly identical ``intrisic'' spectra, but we later apply different noise, reddening and emission line spectra to the intrinsic spectra (described below), producing very different final spectra.  This kind of data augmentation is important for training a convolutional neural network to recognise the underlying stellar population properties despite different added noise, reddening and emission lines.  
Our final training and test samples had 20962
and 5239
galaxies, respectively.

\begin{figure}
    \centering
    \includegraphics[width=\linewidth]{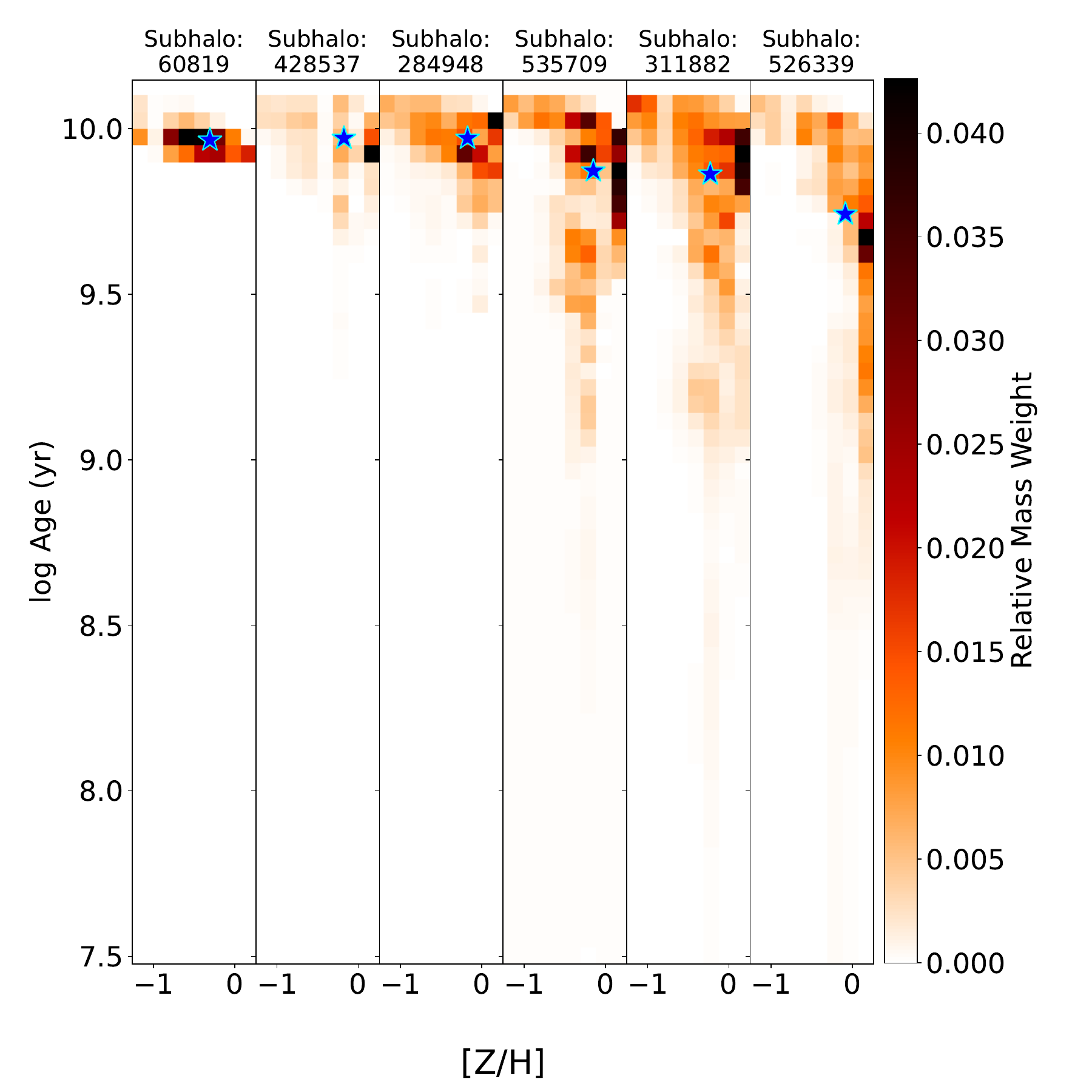}
    \caption{Example star-formation histories (which include metallicity evolution) derived from the ages and metallicities of the star particles of six galaxies drawn from the IllustrisTNG100-1 simulation at snapshot 99.  The ages and metallicities of the star particles of each galaxy are mapped onto an age-metallicity grid, and colour-coded by their mass fraction.  {The star symbols indicate the mass-weighted means of log age and metallicity.}  Note that the actual grid used is an irregularly-spaced grid from the SSPs provided by Vezdekis et al (2016).  The grid here has been regularized and interpolated for illustrative purposes only. }
    \label{TNGsfhs}
\end{figure}

The age,  metallicity and mass of the star particles in each galaxy are provided in the public data release of TNG from which star-formation histories can be constructed.  \fig{TNGsfhs} shows examples of the SFHs (including the metallicity evolution) constructed from the star particles in TNG, mapped onto an age-metallicity grid, and colour-coded by the mass fraction within the grid cells.  The grid cells correspond to 53 stellar ages from 0.03 to 13.5 Gyr, and 8 metallicities [Z/H] (logarithmic metallicity relative to solar) ranging from -1.26 to 0.26, which are covered by the E-MILES simple stellar populations (SSPs) \citep{Vazdekis2016}.  {Specifically, we used the SSPs that assume the \cite{Kroupa2001} universal IMF, \cite{Pietrinferni2004} (BaSTI) isochrones and ``base'' abundances (i.e., following the abundance pattern of the Milky Way).} From these maps we constructed a ``noiseless" spectrum for each galaxy by adding the spectra of the SSPs, weighting the light from each SSP by the mass fraction in the corresponding cell.  {This method is similar in spirit to that of \cite{Nanni2023}, who created mock MaNGA spectra by assigning an SSP to each star particle.}

\begin{figure}
    \centering
    \includegraphics[width=\linewidth]{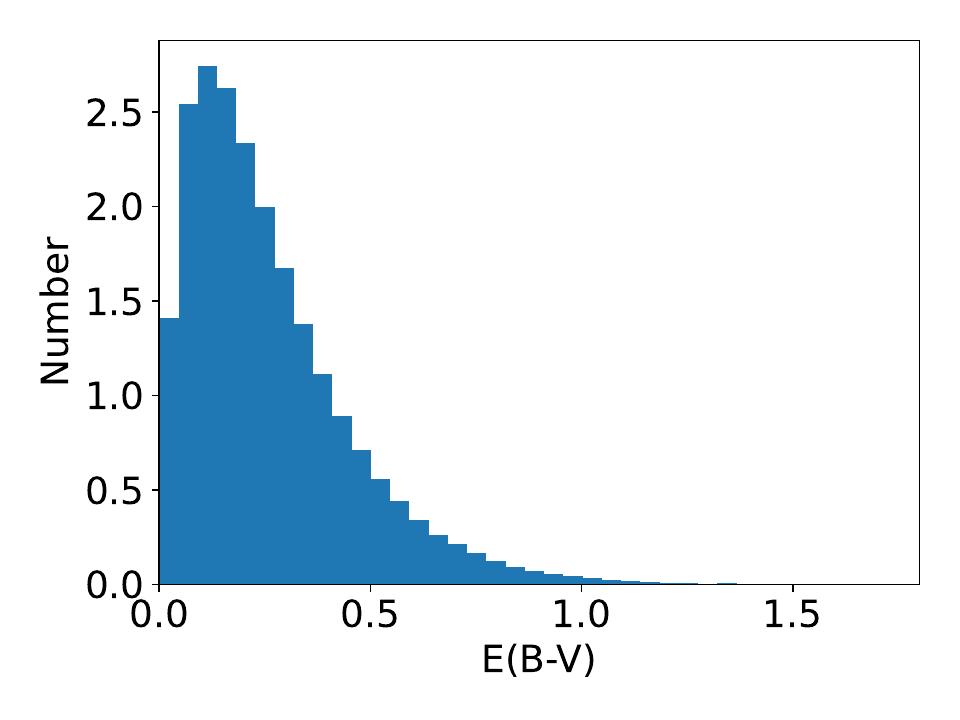}
    \caption{The distribution of internal stellar colour excess E(B-V) values that are applied to our suite of mock spectra.}
    \label{ebvs}
\end{figure}

We redden each spectrum using the \cite{Calzetti2000} extinction curve assuming a uniform foreground screen and random values of the colour excess E(B-V) following the gamma distribution shown in \fig{ebvs}.  This distribution approximates the measured distribution of E(B-V) in the SDSS (as measured by pPXF), but since the effects of reddening on optical spectra are degenerate with some stellar population properties, we did not explicitly draw from any measured distribution of E(B-V) derived from observed spectra.

The second requirement for any set of mock spectra is that their noise properties closely mimic observed spectra.  We chose to mimic spectra from the Sloan Digital Sky Survey (SDSS - DR14).  Common practice in constructing mock spectra is to add Gaussian noise, corresponding to typical noise levels in the observations.  Some (e.g., \citealp{CidFernandas2005,Nanni2023}) add wavelength-dependent levels of noise, according to the wavelength dependence seen in observed spectra.  However, real noise is often not Gaussian. For example, observed spectra (at least from the ground) almost always suffer from sky residuals, particularly at longer wavelengths.  

Furthermore, emission lines from H II regions and/or AGN (including broad lines) contaminate many of the absorption features that are useful markers of stellar population properties (e.g., the Balmer absorption lines).  H II regions can suffer from their own internal reddening that differs from the average reddening of starlight.  Realistic spectra should include realistic emission lines as well as realistic noise.

\begin{figure}
    \centering
    \includegraphics[width=\linewidth]{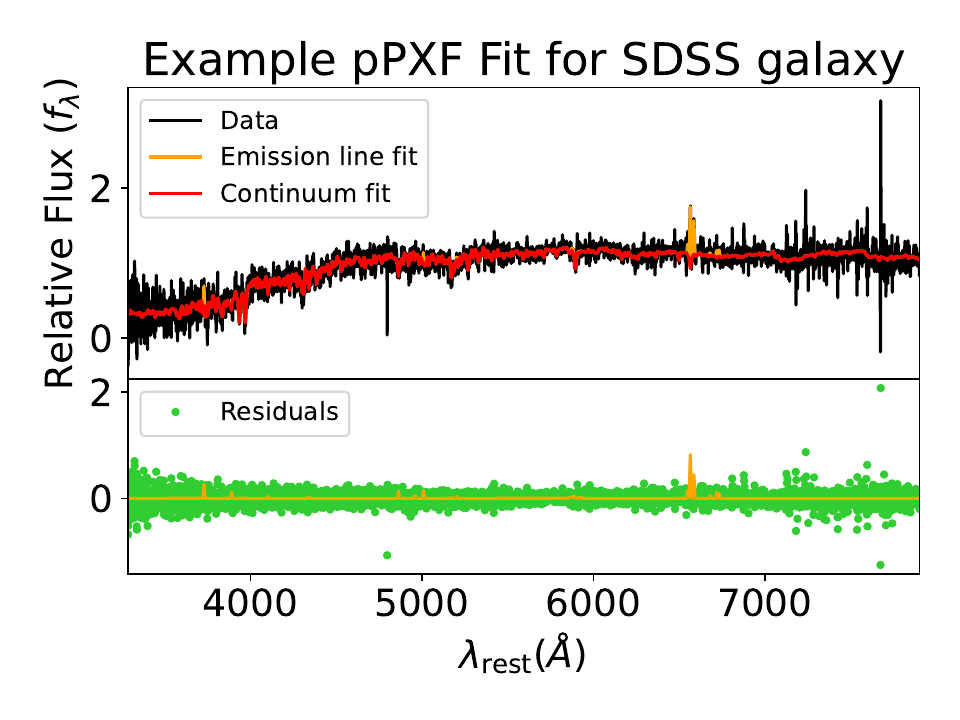}
    \caption{An example of a pPXF fit of the SDSS galaxy with the PlateID of 2573, FiberID of 0262, and observed on the MJD of 54061.  The residuals (green) and the emission lines (orange) in the bottom panel are added to the mock spectra with an unrelated stellar continuum.}
    \label{sdss}
\end{figure}

In order to achieve these levels of realism, we extracted the noise and emission line spectra directly from real galaxies in the SDSS by using one of the conventional spectrum fitting codes (pPXF) to fit both the continuum and emission lines of the spectra.  We chose pPXF since it had the fastest run time of the codes we tested and because it fits the continuum and emission lines simultaneously.  From these pPXF fits on the SDSS, we extracted the residuals, the emission line fits, the fitted redshift and stellar velocity dispersion, but otherwise discarded the stellar component.  An example of an SDSS spectrum fit with pPXF is shown in \fig{sdss}.  The residuals and emission line fits are shown in the bottom panel, and it is these that are retained to be added later to the TNG-generated spectra.
For any given TNG-generated mock spectrum, we randomly selected an SDSS galaxy, and broadened the mock spectrum using the instrumental line-spread-function and velocity dispersion from that SDSS galaxy.  We then shifted the spectrum to the measured redshift of the same SDSS galaxy and added its best-fit emission lines to the mock spectrum.  Note that even if the emission lines were a poor fit, for example due to non-Gaussianity, or the presence of broad line AGN, the fact that we have also added the residuals ensures that the entire line profile is realistic.  An alternative method to extract realistic noise from real spectra is to use an autoencoder (e.g., \citealp{Teimoorinia2022}), but the emission line extraction would have to be done separately.  

\begin{figure*}
   \centering
   \includegraphics[width=\linewidth]{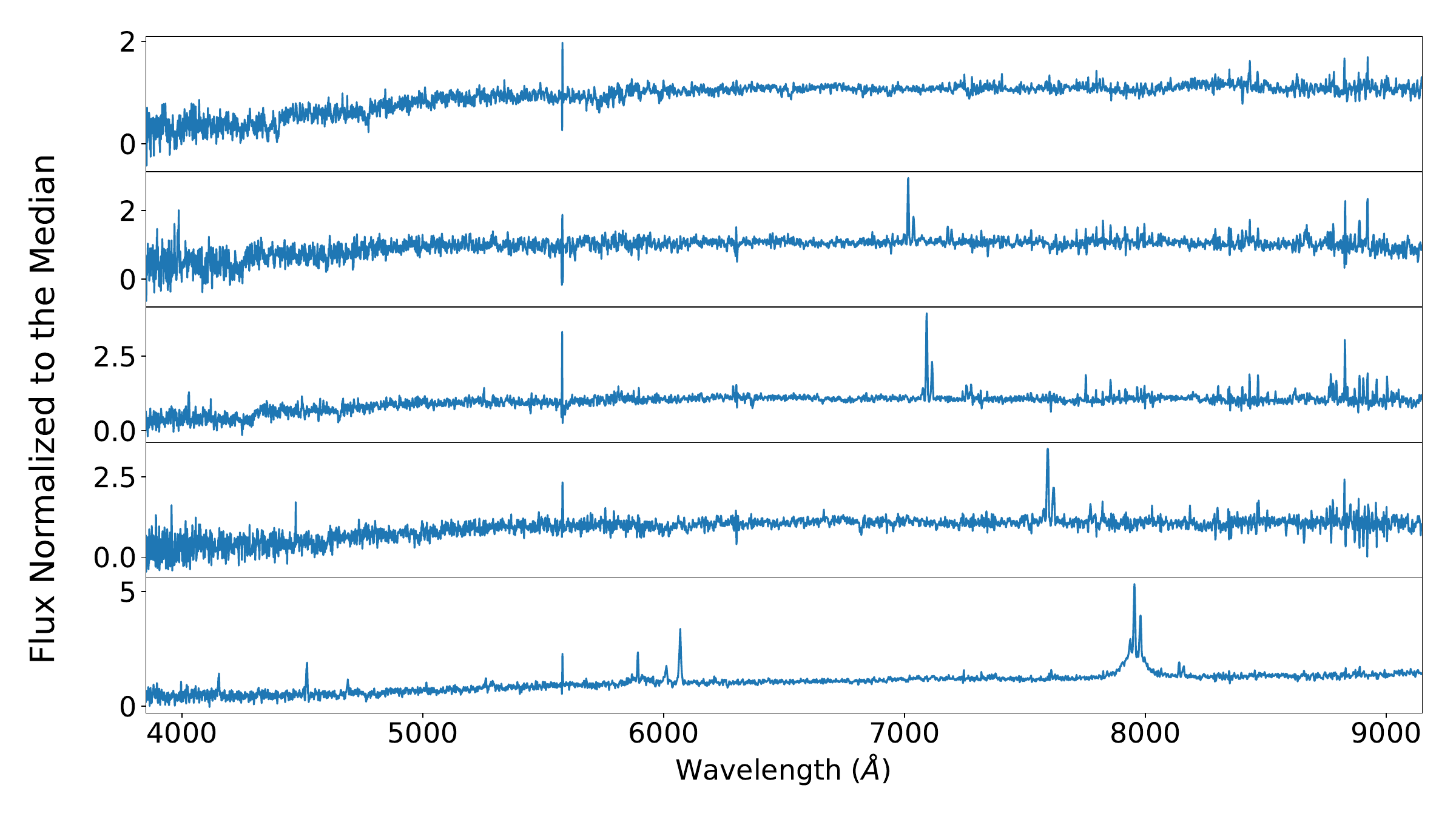}
   \caption{Examples of our final mock spectra created by convolving the E-MILES SSP templates with SFHs from the TNG100 simulation at different redshifts.  These spectra include realistic noise, including sky residuals, emission lines, and includes an example with broad emission from an AGN (bottom panel).}
   \label{examplesim}
\end{figure*}

One effect of the above procedure of extracting emission lines from the SDSS is that the reddening and velocity dispersion of the emission lines is independent of the reddening and velocity dispersion of the stellar population.  This spectrum now represents the ``intrinsic" noiseless spectrum of the galaxy before passing through the Milky Way foreground.  We applied a foreground extinction using the position of the selected SDSS galaxy, the maps of \cite{Schlegel1998}, and the \cite{Fitzpatrick1999} extinction curve.  Lastly we added the residuals from the pPXF fit of the same SDSS galaxy to represent the noise.  Note that all wavelengths from the E-MILES templates were converted to vacuum values before the addition of the SDSS residuals and emission spectra, which are given in the vacuum.  The spectra are sampled in 4544 logarithmically spaced wavelength bins from 3850-9150 {\AA} in the observed frame.  This fits within the typical wavelength range of most SDSS spectra to ensure that the residual spectra adequately cover the range of the mock spectra.  Examples of our final mock spectra are shown in \fig{examplesim}, which includes an example of a broad-line AGN spectrum.

To mimic the information available in real SDSS spectra, in the header of each mock spectrum file we include the redshift, velocity dispersion, original noise spectrum (not the pPXF residuals), foreground E(B-V), and line-spread-function FWHM from the real SDSS galaxy that provided the realism of the mock spectrum.  Thus we have made it ``easy" for the spectrum fitting codes by providing the exact values of these quantities for their initial guesses (for the redshift, velocity dispersion), so that only their ability to determine the properties (age, metallicity, reddening, $\MLr$) of the underlying stellar population is tested.

The procedure above ensures that our suite of mock spectra shares the same distribution of redshifts, velocity dispersion (of both the stars and gas), emission spectra, noise profiles and even foreground extinction as our SDSS sample, so a description of the sample is in order.  The SDSS sample was selected to have $z< 0.3$, classified as a ``science primary" and not a star or a sky spectrum as denoted by the ``specClass" flag (709537 galaxies).  Furthermore, the pPXF fit must have succeeded (``status" value of greater than 0), achieved a $\chi^2 < 3$ and measured a stellar velocity dispersion less than 1000 km/s (690790 galaxies).  Lastly, some SDSS spectra are truncated with starting wavelengths in excess of 5000 \AA, or longest wavelengths shorter than 6500 \AA, making them unsuitable for extracting residuals.  Requiring that the SDSS spectra cover a wavelength range of 3850-9150 {\AA} resulted in our final sample of 613411 galaxies.  It is from this sample that redshifts, velocity dispersions, emission lines and residuals are applied to our mock spectra.

Although our mock spectra share the same emission lines as SDSS galaxies, since they are drawn randomly from the SDSS sample, the emission lines are completely uncorrelated with the SFHs that created the underlying continuum spectra.  Therefore these spectra are ideal for training a neural network to identify stellar population properties that are unrelated to emission lines, but are probably unsuitable for testing fitting codes, such as FADO \citep{Gomes2017}, that use the emission lines to constrain the SFH.

Lastly, recall that our TNG sample included only 26201 galaxies.  We have found that the StarNet CNN required a much larger sample to train on, so we augmented our sample to 250000 spectra by recycling the original 26201 SFHs many times, but applying different redshifts, dispersions, emission lines and noise from different randomly drawn SDSS galaxies. 200000 of these were produced from our original training sample (80\% of the total, or 20962) and were set aside to train StarNet. The other 50000 were produced from our original test sample (20\%, or 5239) and were used to test StarNet as well as the conventional fitting codes. This procedure ensured that no SFH appeared in both sets.

\section{The spectrum fitting Codes}

\subsection{Commonalities and Differences in Setup Conditions}
\label{setup}

Here we summarize the algorithms of the spectrum fitting codes that we compare and the settings that we used to run them.  Since our goal is to compare them on as equal a footing as possible, there were a number of conditions and preparatory steps that were held in common before running the codes.  

The same mock spectra generated from TNG SFHs were used for all the codes, and they were transformed to the rest-frame using the redshift information included in the header before feeding them to the codes since this was required by pPXF and {\sc starlight}.  Furthermore, we set each code to use the same reddening law \citep{Calzetti2000} to fit $A_V$ or E(B-V).  (This was the same law that we used when constructing the mock spectra.)

We also fitted the mock spectra with a subset of the same E-MILES SSP templates used to create the spectra.  Since many of the SSP templates differ from each other by an amount smaller than the typical errors, standard practice in full spectrum fitting is not to use the full suite of available SSP templates, but to use a coarse grid in stellar age for the oldest templates, and finer binning for the youngest.  Our grid of chosen templates includes 16 bins of stellar age: (0.03, 0.05, 0.07, 0.09, 0.2, 0.4, 0.7, 1, 1.5, 2, 3, 5, 7, 9, 11 and 13.5 Gyr), and 8 bins of metallicity ([Z/H] = -1.26, -0.96, -0.66, -0.35, -0.25,  0.06,  0.15 and 0.26), making a total of 128 templates.  So, although a finer grid of 53x8 E-MILES templates was used to construct the spectra, we use a coarser grid to fit them.  Thus, for all codes the time- and metallicity-resolutions of the SFH's are limited by our coarser grid.

Note that in the current version of Firefly (v1.0.1), {the user can choose between various flavours of the Maraston models for different input libraries including MILES, ELODIE, STELIB, MARCS, and MaStar (Maraston et al. 2020).  We added a few lines to Firefly to read the E-MILES templates instead.}  However we also repeated the Firefly runs to fit MaStar templates to mock spectra created with MaStar templates, and found that our results were not significantly different.

The StarNet CNN uses information in the spectrum itself to compute stellar population properties (as described in \secref{starnet}), and hence no SSP templates are used for StarNet.

While desiring a fair comparison between the codes, we also wished to compare them as much as possible on the basis of their ``out-of-the-box" features, \ie, those requiring minimal pre-processing on the the part of the user.  Therefore, some built-in features were by necessity different between the codes.  

A major feature {\it not} kept in common between the codes was the treatment of emission lines.  Both pPXF and pyPipe3D fit emission lines as part of the spectrum fitting process, but Firefly and {\sc starlight} do not.  Therefore, for the latter two codes, we masked the regions of the following emission lines, while for the former two we fit them explicitly: 
\begin{itemize}
    \item {[OII]}3726, 3729
    \item {[OIII]}4959, 5007
    \item {[NII]}6548, 6583
    \item {[SII]}6717, 6731
    \item Balmer lines from H$\alpha$ to H$\delta$
\end{itemize}
These were chosen since they are the brightest nebular lines, and all other lines are usually insignificant.  {pPXF includes an example of how to add emission lines to the fit (\texttt{ppxf\_utils.py}).}  It was simple to comment out the extra emission lines that are not in the above list.  Sky residuals were not masked or fitted for any of the codes.  

As described in \secref{simulatedspectra}, some of the mock spectra contain broad-line AGN emission in the Balmer lines because the emission lines are drawn from the SDSS.  None of the codes are designed to deal with them properly: Firefly and {\sc starlight} require line masking (we chose masks 9 \AA or 400 km/s) which do not adequately cover broad lines.  pPXF fits emission lines, but imposes a maximum velocity dispersion of only 1000 km/s. {This can be adjusted by the user, but we have kept the default, and} have only fit a single component for simplicity.  pyPipe3D's default configuration fits a maximum velocity dispersion of 5.5 \AA~ for the H$\alpha$ line and 7.5 \AA~ for H$\beta$, also insufficient for covering the AGN.  Therefore we expected that the stellar population ages to be poorly determined for these spectra for the non-DL codes.

Note that StarNet, as all CNNs, performs computations on uniform grids of data, and hence is not able to handle masked data, such as masked emission lines at different redshifts.  Therefore, emission line masking is not performed for StarNet. 

A less significant difference between the codes is the treatment of the foreground extinction.  Firefly reddens its templates with the foreground value before fitting (using the \citealp{Fitzpatrick1999} parametrization) and so no de-reddening beforehand is necessary.  However the other codes required de-reddening of the mock spectra for the foreground as a pre-processing step (also using the \citealp{Fitzpatrick1999} curve).  We also performed a de-reddening of the foreground extinction before running the spectra through StarNet.

We also compared the run time for each of these codes.  Run time will of course depend on a number of factors including the number of stellar templates used (which we made the same for all the codes), whether emission lines are fitted (2 out of the 4 conventional codes), how many pixels are masked (differs between the codes), and of course, the CPUs involved.  We used the generous national resources of Digital Resources Alliance of Canada (formerly Compute Canada), specifically the Cedar heterogeneous compute cluster.  We ran the codes in an ``embarrassingly parallel" manner (which refers to the parallelization scheme where there is no communication between processes) and report the average runtime of each code for a single spectrum on a single CPU in \tab{comparisontable}.

Our goal was to test the codes' ability to recover the following four properties of the population: the mean mass-weighted logarithm of the stellar age, the mean mass-weighted stellar metallicity, the $r$-band stellar mass-to-light ratio, and the colour excess E(B-V).  {As is common in the literature, the mass-weights for the age and metallicity refer to the total stellar mass ever formed (i.e., the integral of the SFH), while the mass-to-light ratio refers to the current living stars only.}  The non-DL fitting codes compute much more information, such as the full SFH, and some estimate the errors on the outputted parameters.  However the StarNet CNN needs to be trained to recover specific parameters, and we chose the aforementioned four.  Future studies will test and compare the recovery of the full SFH.  

In the next subsections, we summarize the algorithms of each code (in alphabetical order) as well as the specific preprocessing steps we performed to run each code.  \tab{comparisontable} summarizes the salient points of each code.

\begin{table*}
    \centering
    \renewcommand{\arraystretch}{1.5}
    \begin{tabular}{p{2.1cm}p{2.55cm}p{2.85cm}p{2.65cm}p{2.55cm}p{2.65cm}}
    \hline
     & \colh{Firefly} & \colh{pPXF} & \colh{pyPipe3D} & \colh{STARLIGHT} & \colh{StarNet} \\
     \hline 
    {\bf Version} &  v1.0.1$^{1}$ & v8.1.0$^{2}$ & v1.1.5 & v04 & N/A$^{3}$ \\
    {\bf Language} & Python & Python & Python & Fortran & Python \\
    {\bf Runtime (per spectrum)} & 43s & 11s & 97s & 99s & 0.0004s (+9 min one-time training) \\
    {\bf Emission lines} & masked (9 \AA) & simultaneous fit & separate fit & masked (400 km/s) & not fitted \\
    {\bf Regularization} & No & Yes & No & No & No \\
    {\bf Algorithm} & 
    Sum of SFHs weighted by $\chi^2$ likelihoods & 
    {Levenberg-Marquardt and {\sc CapFit} simultaneously fit all parameters} & 
    Step 1: fit $z$, $\sigma$ and $A_V$.  Step 2: fit emission lines.  Step 3: MC fitting of SFHs & MCMC to find minimum of SFH parameter space & 
    CNN: 2 convolutional layers, 1 max pooling layer, 3 fully connected layers \\
    \hline
    
    \end{tabular}
    \caption{Summary of spectrum fitting codes.  Notes: $^{1}$Firefly was modified to use the E-MILES templates, and fix 2 bugs (see \secref{firefly}).  $^{2}$pPXF was modified to only include the emission lines listed in \secref{setup}.  $^{3}$StarNet is a simple script that was kindly provided by one of the co-authors of Fabbro et al. (2018). }
    \label{comparisontable}
\end{table*}

\subsection{Firefly}
\label{firefly}

Firefly (\citealp{Wilkinson2017} and {updated in \citealp{Neumann2022}}) is a $\chi^2$ minimization code that attempts to broadly explore the $\chi^2$ parameter space of SFHs, while avoiding unnecessarily complex solutions that prolong computation time.  The basic algorithm is to test the $\chi^2$ values between the model and data linear flux, {combining increasingly complex SFHs that consist of a linear combination of equally weighted SSPs}.  The SFHs that do not decrease $\chi^2$ compared to simpler SFHs (\ie, those consisting of fewer SSPs) are discarded.  After obtaining a set of acceptable SFHs (usually numbering in the thousands), the likelihood of each SFH is computed from the chi-squared probability distribution.  The final solution is the sum of each SFH, weighted by their likelihoods.  {Firefly performs an initial run of this algorithm to produce an unreddened best-fit spectrum. A reddening spectrum is then computed from the difference between data and the best-fit, which is then applied to all templates. Then the fitting is performed a second time on the original spectrum with the reddened templates. A dust reddening law like \cite{Calzetti2000} is fit separately in order to derive an E(B-V) value, which is however not used in the fitting procedure.}

Before fitting, if the user chooses, Firefly reddens the SSPs with a foreground extinction, using the sky coordinates provided by the user and the dust maps of \cite{Schlegel1998} and the dust law of \cite{Fitzpatrick1999}.  We took advantage of this feature by providing the foreground E(B-V) values we used in the construction of the mock spectra.

Firefly does not explicitly fit emission lines.  As suggested by \cite{Wilkinson2017}, one could provide emission lines as templates, or fit and subtract and the emission lines separately before using Firefly.  However, in the interest of minimizing the pre-processing needed by the user, we chose to use Firefly's options to mask emission lines.  Firefly allows the user to choose the wavelength window around the lines to mask (we chose 9\AA), as well as which lines out of a list of 18 lines and doublets.  We masked the lines listed in \secref{setup}.

We added a section to Firefly's code (to the \texttt{get\_model} function in \texttt{firefly\_models.py}) in order to use the E-MILES SSP templates, following the same format as the reading and preprocessing of the MaStar templates.  (We also ran Firefly fitting MaSTAR templates to mock spectra created with MaStar templates and found similar results.)  After converting the wavelengths to vacuum values, we sampled the templates with logarithmic binning (to mimic our input spectra).  We used the default option in Firefly to degrade the templates to the resolution of the galaxy velocity dispersion given the instrumental resolution, and then followed Firefly's procedure of reddening the templates using the foreground E(B-V).  Firefly later rebins the templates to the same binning as the input spectrum, and then normalizes the templates in order to supply SSP weights in both light and mass.

While testing, we discovered and fixed two bugs and reported them on the Firefly github page (and by e-mail to the Firefly authors).  The first bug is that the emission line masking assumes air wavelengths regardless of the user's specification of whether the data are in air or vacuum wavelengths.  This had the effect of not completely masking the emission lines for small masking windows (such as our window of 9\AA) when the data are in vacuum wavelengths.  The second bug is that, although we kept the default \texttt{max\_ebv} of 1.5 (\ie, allow a maximum best-fit E(B-V) of 1.5), the actual maximum E(B-V) was 0.67 so that highly reddened populations had large errors in their stellar population properties.  However, this bug did not have a significant effect on the aggregate of Firefly's results since most of the best-fit E(B-V) values were well below 0.67.  These two bugs remained unaddressed at the time of paper submission. What we present in this paper are the results of Firefly after applying these bug fixes. {A new version of Firefly has been published on Github in the meantime, in which both these bugs have been fixed.}

The output of Firefly consists of weights (both light and mass) on the templates that when combined produce the best-fit of the spectrum.  Firefly outputs both the light-weighted and mass-weighted stellar age and metallicity as well as the E(B-V) of the stars.  We used the mass weights to compute $\MLr$.

The average runtime of Firefly with our chosen setup on Cedar was 43 seconds per spectrum on a single CPU.

\begin{figure*}
    \centering
    \includegraphics[width=\linewidth]{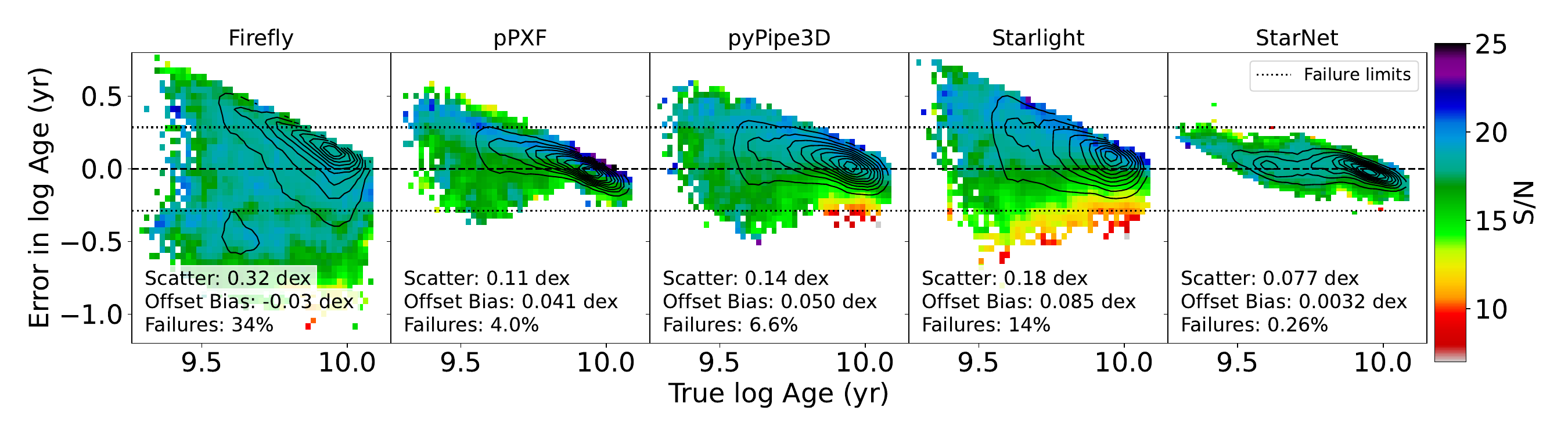}
    \includegraphics[width=\linewidth]{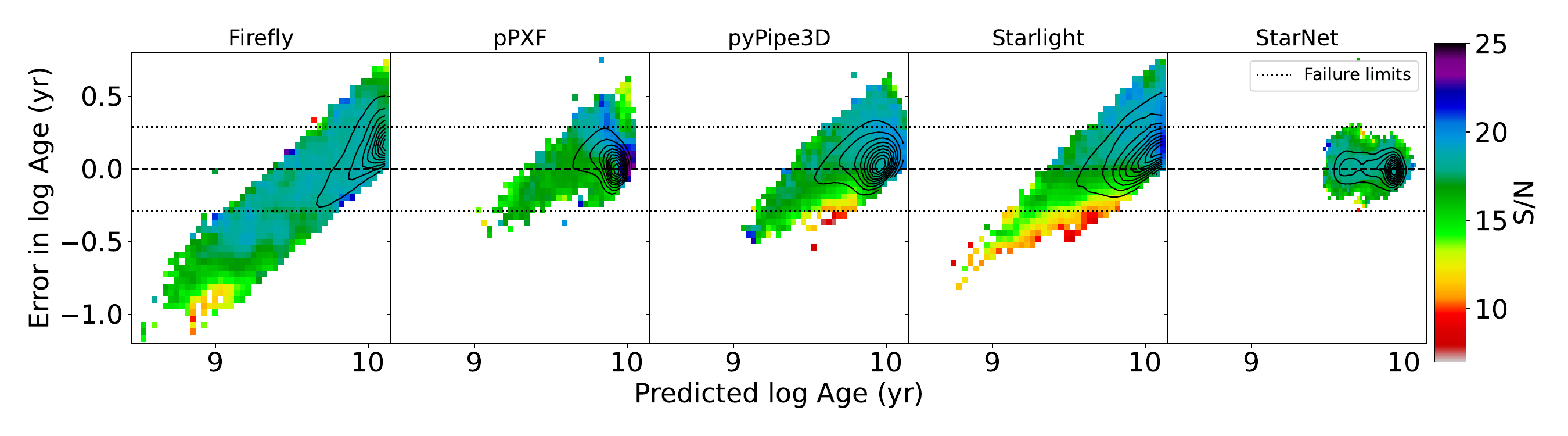}
    \caption{The recovery of stellar age for the 5 indicated codes.  The $y$-axis is the predicted log Age minus the true log Age, while the $x$-axis is the true log Age (top) and the predicted log Age (bottom).  The contours represent the number density of points in the plot, while the colour scale is the mean S/N. ``Failures" are defined in \secref{recovery}.}
    \label{lage}
\end{figure*}

\begin{figure*}
    \centering
    \includegraphics[width=\linewidth]{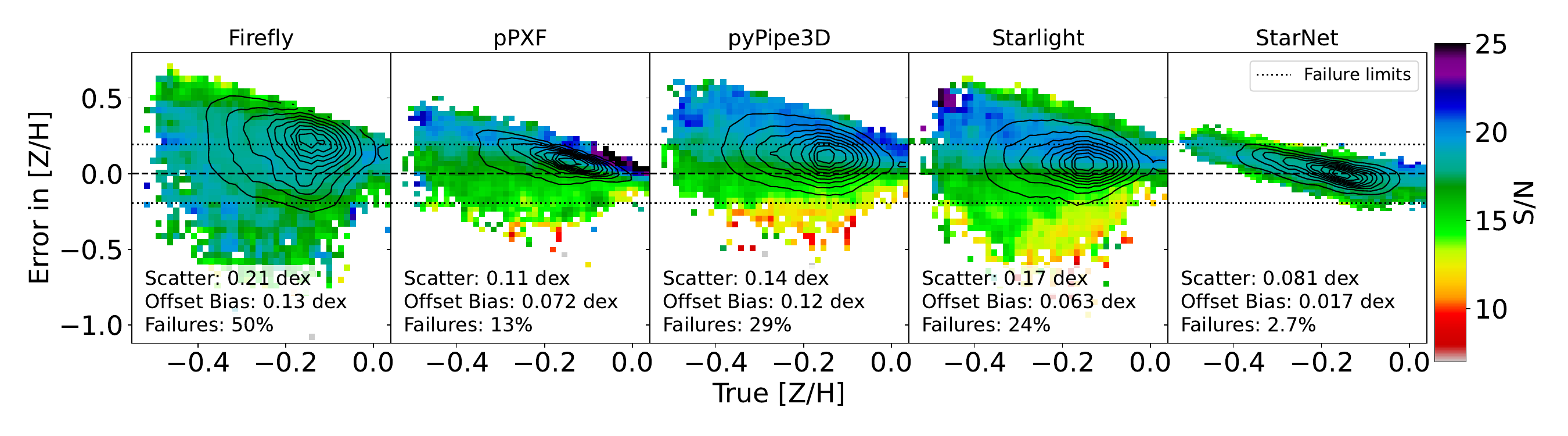}
    \includegraphics[width=\linewidth]{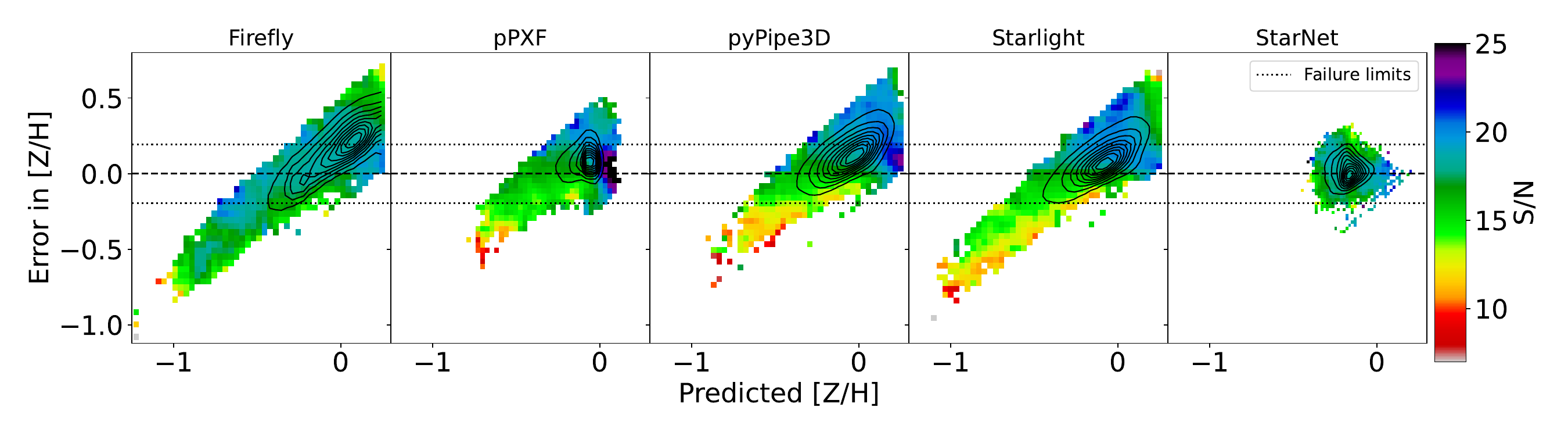}
    \caption{The recovery of stellar metallicity for the five indicated codes.  The $y$-axis is the predicted [Z/H] minus the true [Z/H], while the $x$-axis is the true [Z/H] (top) and the predicted [Z/H] (bottom).  The contours represent the number density of points in the plot, while the colour scale is the mean S/N. ``Failures" are defined in \secref{recovery}.}
    \label{metal}
\end{figure*}

\begin{figure*}
    \centering
    \includegraphics[width=\linewidth]{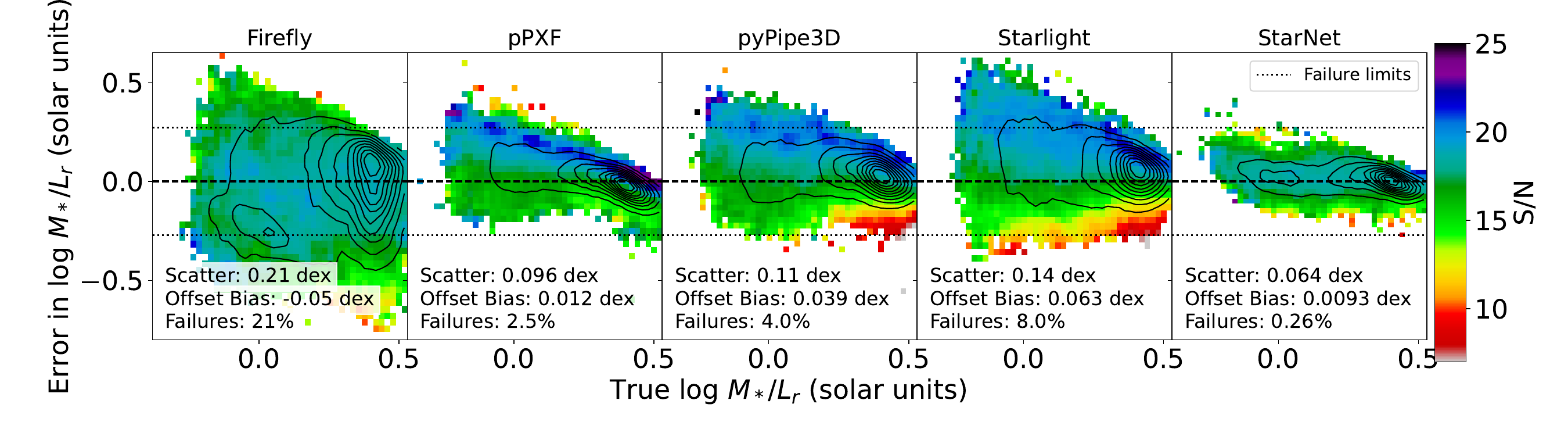}
    \includegraphics[width=\linewidth]{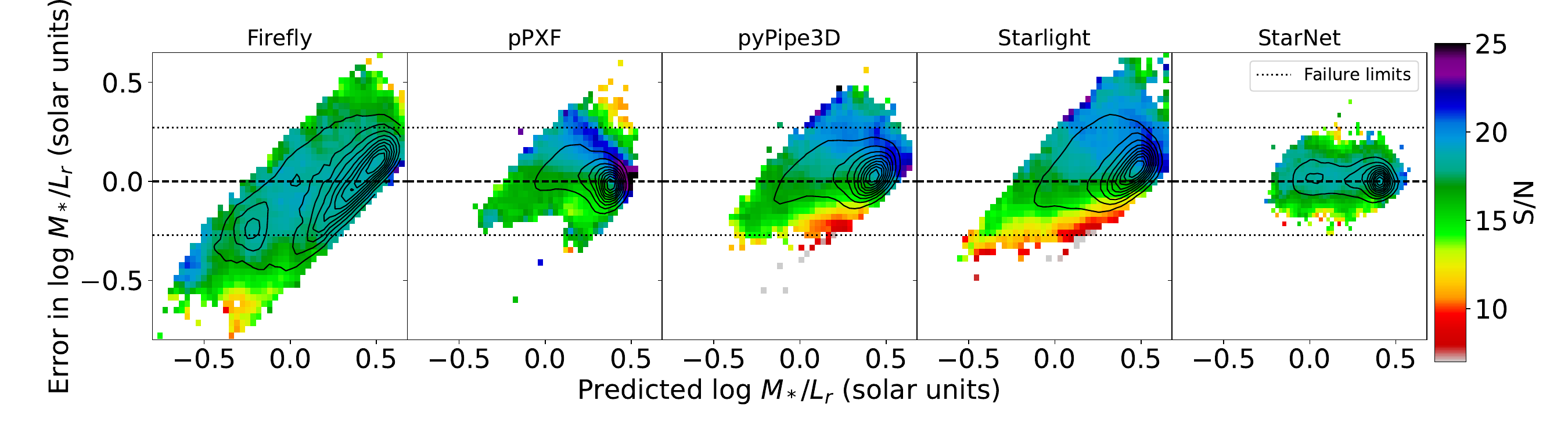}
    \caption{The recovery of stellar $\MLr$ for the five indicated codes.  The $y$-axis is the measured log $\MLr$ minus the true log $\MLr$, while the $x$-axis is the true log $\MLr$ (top) and the predicted log $\MLr$ (bottom).  The contours represent the number density of points in the plot, while the colour scale is the mean S/N. ``Failures" are defined in \secref{recovery}.}
    \label{lML}
\end{figure*}

\begin{figure*}
    \centering
    \includegraphics[width=\linewidth]{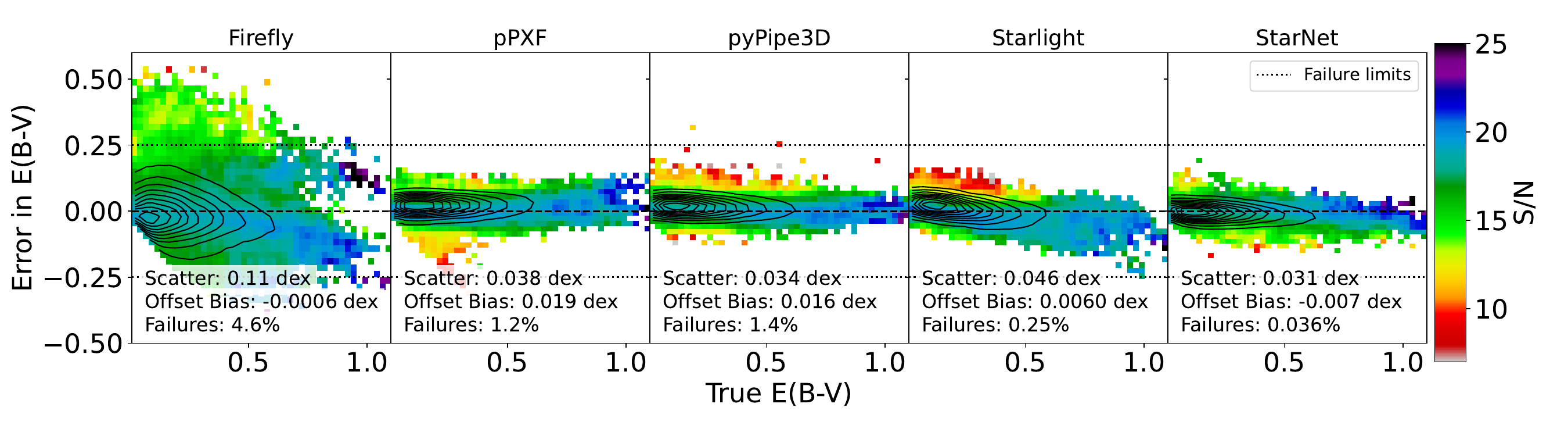}
    \includegraphics[width=\linewidth]{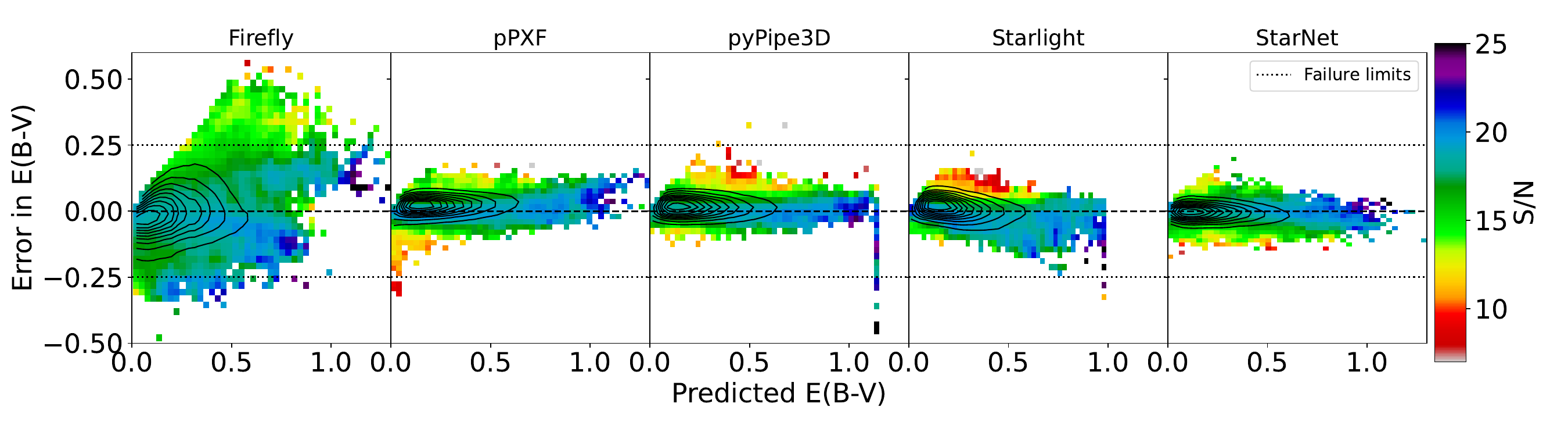}
    \caption{The recovery of the colour excess (of the stars) for the five indicated codes.  The $y$-axis is the measured E(B-V) minus the true E(B-V), while the $x$-axis is the true E(B-V) (top) and the predicted E(B-V) (bottom).  The contours represent the number density of points in the plot, while the colour scale is the mean S/N. ``Failures" are defined in \secref{recovery}.}
    \label{ebv}
\end{figure*}

\subsection{pPXF}

pPXF \citep{Cappellari2004,Cappellari2017} is another $\chi^2$ minimization method that fits the line-of-sight velocity distribution (LOSVD), stellar population weights, dust reddening and emission lines using a set of stellar and emission line templates.  
{The authors of pPXF recognized that, for a fixed set of kinematic parameters, the optimal  template weights can be obtained by solving a quadratic optimization problem, which has an efficient and exact solution (as was first noted and exploited by \citealp{Faber1972}).  
Therefore, pPXF optimizes the linear (stellar population weights) and nonlinear (kinematics, reddening, etc.) parameters of the spectrum fitting problem simultaneously, but by using different methods. The linear parameters are solved by quadratic programming. The nonlinear parameters are optimized by a novel algorithm (\textsc{CapFit} since pPXF v6.5 in 2017 - \citealp{Cappellari2023}), which combines the well-known Levenberg-Marquardt non-linear least-squares algorithm (\eg, \citealp{Press2007}) with the sequential quadratic programming method that allows for more complex parameter constraints. The fact that pPXF exploits the quadratic nature of the template weights fitting sub-problem is likely the reason why pPXF is significantly faster than the other non-DL codes we have tested.}

The ``penalized" part of pPXF's name refers to the penalty that is applied to the $\chi^2$ merit function, which penalizes non-Gaussian forms of the LOSVD.  The penalty is applied in such a way as to preserve the quadratic nature of the problem.  

pPXF has the option of regularizing the SFH by finding the smoothest distribution of SSP weights that is allowed by the data, given a chosen regularization error.  The regularization error is a user-defined parameter that determines the maximum smoothness the solution should seek.  \cite{Cappellari2017} states that the regularized solution can be interpreted in a Bayesian sense to be the most likely solution for the SSP weights given a prior on the amplitude of fluctuations in the SFH.  pPXF achieves regularization of the SSP weights in such a way as to preserve the quadratic nature of the $\chi^2$ minimization.  Following the suggested procedure in the documentation, we normalized the templates and input spectra and used a regularization error of 0.01 (\texttt{regul=100}).  {We did test other values of \texttt{regul}, and found that pPXF's parameter recovery improved with higher values of \texttt{regul}, but improvements were minimal beyond \texttt{regul=100}.}

pPXF also has the option to fit multiplicative and additive polynomials in order to correct the continuum shapes and reduce $\chi^2$ further.  However we did not enable either of these options.

In order to use the E-MILES templates, we converted the template wavelengths to vacuum values, convolved the templates to the same instrumental resolution of the test spectrum, and resampled the templates to be spaced logarithmically in wavelength.  
pPXF also requires that the spectra be transformed to the rest-frame, which we did using the redshift information that was saved in the headers during the creation of the spectra.  

As mentioned above, pPXF performs emission line fitting simultaneously with the stellar population weights and LOSVD, with the kinematic properties of the gas and stars being allowed to differ.  The code package includes a function in \texttt{ppxf\_util.py} to provide emission lines as separate templates, {to be modified by the user as needed}.  The list of emission lines includes the strong lines listed in \secref{setup}, as well as the Balmer lines from H$\epsilon$ to H10, and [OI]6300/6364.  It was not difficult to comment out the extraneous emission lines in order to be consistent with the emission lines we used in the other codes.  

The output of pPXF consists of mass weights on the templates that when combined produce the best-fit of the spectrum, as well as E(B-V).  We used these weights to compute $\MLr$, and the mass-weighted stellar age and metallicity.

The average runtime of pPXF with our chosen setup on Cedar was 11 seconds per spectrum on a single CPU.

\subsection{pyPipe3D}

pyPipe3D (\citealp{Lacerda2022}) is a Python version of Pipe3D \citep{Sanchez2016}, which was originally written in Perl and C.  The philosophy of pyPipe3D was to make the code more modular and accessible to the astronomical community, while preserving the original algorithms.  pyPipe3D performs its spectrum fitting in three steps.  

First, pyPipe3D fits the redshift, stellar velocity dispersion and extinction $A_V$, exploring the $\chi^2$ parameter space one parameter at a time.  This step involves the use of a skeleton set of SSPs to speed up the process.  The public version of the code uses three SSPs with various ages and metallicities. For this step, we chose three E-MILES templates with similar ages and metallicities as those used in the default set (ages: 0.07, 1, 12.5 Gyr, [M/H]: -0.66, -0.35, 0.26 dex, respectively).  
At this stage, emission lines are masked, as defined by the user, and we chose the same list of emission lines above.

Second, pyPipe3D subtracts the best-fitting combination of the three SSPs in the first step from the observed spectrum, leaving the noise and emission lines.  The algorithm then fits a user-defined list of emission lines, deriving their Gaussian parameters by a combination of a random Monte-Carlo exploration of the parameter space (for robustness) and the Levenberg-Marquardt method for speed.  We used the same list of emission lines in \secref{setup}.  One implication of this algorithm is that the spectral segment underlying the emission lines might be less well determined than in pPXF since only three templates were used to model it.

Third, these fitted emission lines are subtracted from the original spectrum leaving a gas-free stellar continuum. The continuum is fitted as a linear combination of a larger set of SSPs (the same E-MILES set we have used for all the fitting codes), which have been transformed using the redshift, velocity dispersion and $A_V$ found in the first step.  The fitting is a Monte-Carlo exploration of the parameter space of SSP weights that produce a minimum in $\chi^2$.  Several realizations (the number is not specified in \citealp{Lacerda2022}) of the spectrum are created by perturbing the spectrum on levels consistent with the noise and the Monte-Carlo fitting is performed for all realizations.  This generation and fitting of multiple realizations is likely the reason why pyPipe3D is among the slowest of all the codes we tested.  The final solution and error in the fit are the average spectrum and standard deviation over the realizations.  The SFH solution and errors are the average and standard deviations of the weights for each SSP.

The preprocessing required for running pyPipe3D involved transforming the E-MILES templates into vacuum wavelengths, normalizing them at a particular wavelength (we chose 5500.7 \AA) and saving them in the well-documented format that pyPipe3D requires.  Although pyPipe3D does not require it, we also transformed our spectra to the rest frame, since this was done for the other fitting codes (pPXF and {\sc starlight}), or the code itself did the de-redshifting (Firefly).  

The output of pyPipe3D consists of mass- and light-weighted ages and metallicities, $A_V$ (which we converted to E(B-V) using $R_V=4.05$), and the light weights on the templates that when combined produce the best-fit of the spectrum.  We used the weights to compute $\MLr$.

The average runtime of pyPipe3D with our chosen setup on Cedar was 97 seconds per spectrum on a single CPU.

\subsection{{\sc starlight}}

{\sc starlight} \citep{CidFernandas2005} minimizes $\chi^2$ using a mixture of simulated annealing and Metropolis MCMC techniques to explore the parameter space of SFHs.  Whereas Firefly and pyPipe3D fit E(B-V) separately from the stellar population fitting, {\sc starlight} (like pPXF) includes a multiplicative extinction factor as a fitting parameter in the model, assuming an extinction law chosen by the user (we chose \citealp{Calzetti2000}).  The other free parameters are the SSP weights, the systemic velocity and stellar velocity dispersions, although the kinematic components are fit separately from the other parameters (after each annealing loop) for reasons of efficiency.  The Metropolis algorithm explores the whole parameter space but is designed to gravitate towards the region of highest likelihood, thereby avoiding local minima in $\chi^2$.  {\sc starlight} runs in 4 stages: a first fit which finds the general location of the global minimum in $\chi^2$, a sigma-clipping stage for poorly fit data points, a ``burn-in'' stage in which it fine tunes the location of the $\chi^2$ minimum, and a final stage removing SSPs with negligible contribution and fitting again. 

{\sc starlight} can be configured with many options to do with the fitting process, including the Markov chain parameters and clipping parameters.  The package includes example configuration files and modifications corresponding to a ``slow", ``medium" and ``fast" configuration.  We chose the fast configuration.

The preprocessing required to run {\sc starlight} includes de-reddening the foreground, transforming the spectrum into the rest-frame (which we did using the redshift saved in the headers during the creation of the spectra), resampling the spectrum onto a linearly spaced grid (instead of logarithmic) with a spacing of 1 {\AA} as instructed in the documentation, and masking emission lines and sky residuals.  {\sc starlight} does not automatically mask any emission lines, so we had to do this ourselves.  The lines we masked were 400 km/s on either side of the lines listed above. 

Preprocessing on the templates was merely to convert the E-MILES wavelengths to vacuum values.  {\sc starlight} is able to handle the irregular spacing that this conversion produces. 

The output of {\sc starlight} consists of the $A_V$ of the stars (which we converted to E(B-V) using $R_V=4.05$), and the mass- and light-weights on the templates that when combined produce the best-fit of the spectrum.  We used the weights to compute the mass-weighted stellar age, metallicity and $\MLr$.

The average runtime of {\sc starlight} with our chosen setup on Cedar was 99 seconds per spectrum on a single CPU.

\begin{figure*}
    \centering
    \includegraphics[width=\linewidth]{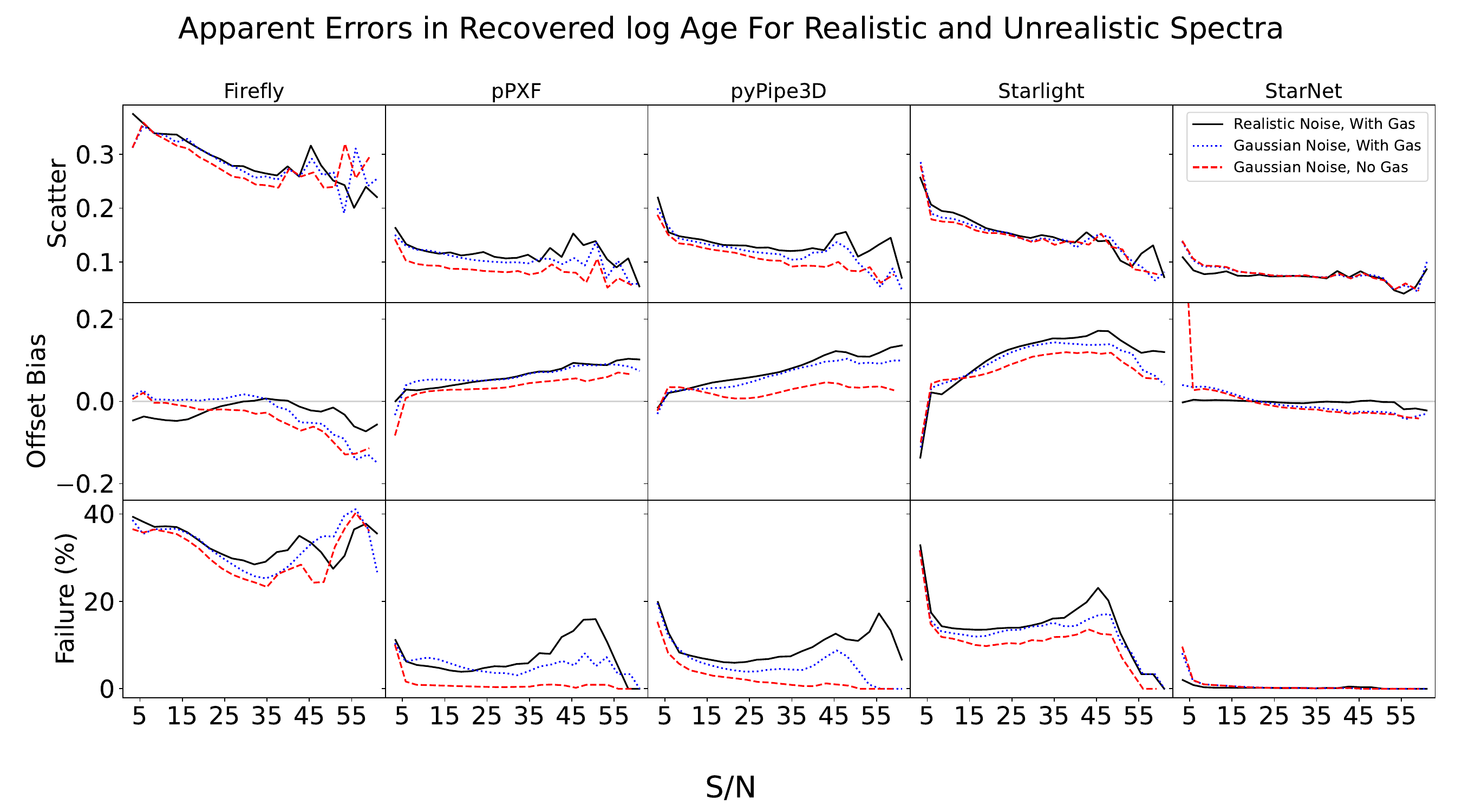}
    \caption{The effect of noise and gas emission lines on the recovered stellar age, namely the scatter in the errors, the mean offsets, and the failure rate as a function of mean S/N in each grid cell for the five tested codes.  Emission lines seem to have a bigger effect on the recovery of stellar age than the type of noise applied (the red dashed curves are more offset from the fiducial solid black curves than the blue dotted curves).}
    \label{errorsGas_lage}
\end{figure*}

\begin{figure*}
    \centering
    \includegraphics[width=\linewidth]{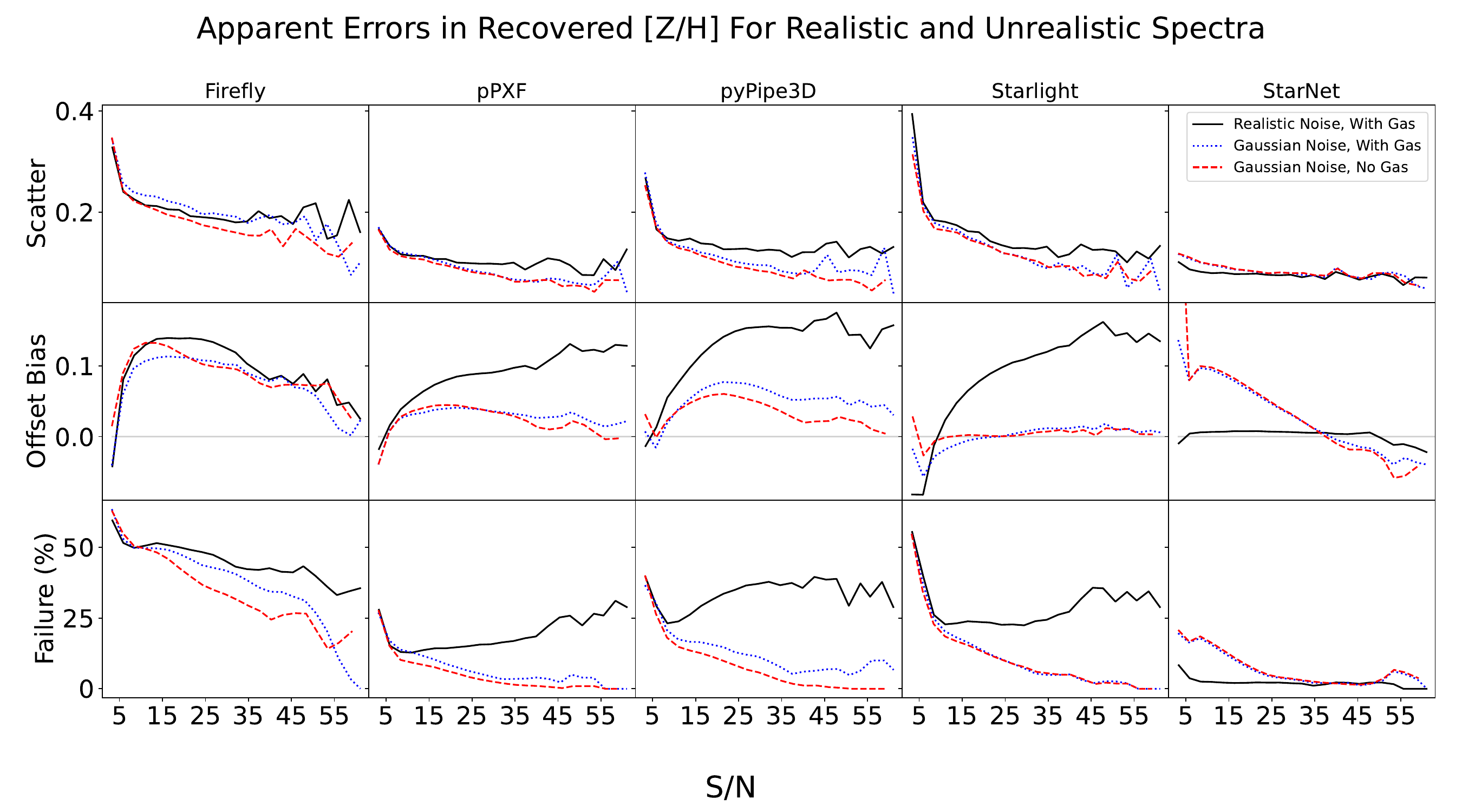}
    \caption{The effect of noise and gas emission lines on the recovered stellar metallicity, namely the scatter in the errors, the mean offsets, and the failure rate as a function of  mean S/N in each grid cell for the five tested codes.  The type of noise seems to have a bigger effect on the recovery of stellar metallicity than the presence of emission lines (the red dashed curves are close to the blue dotted curves, and they are both significantly offset from the fiducial solid black curves).}
    \label{errorsGas_metal}
\end{figure*}

\begin{figure*}
    \centering
    \includegraphics[width=\linewidth]{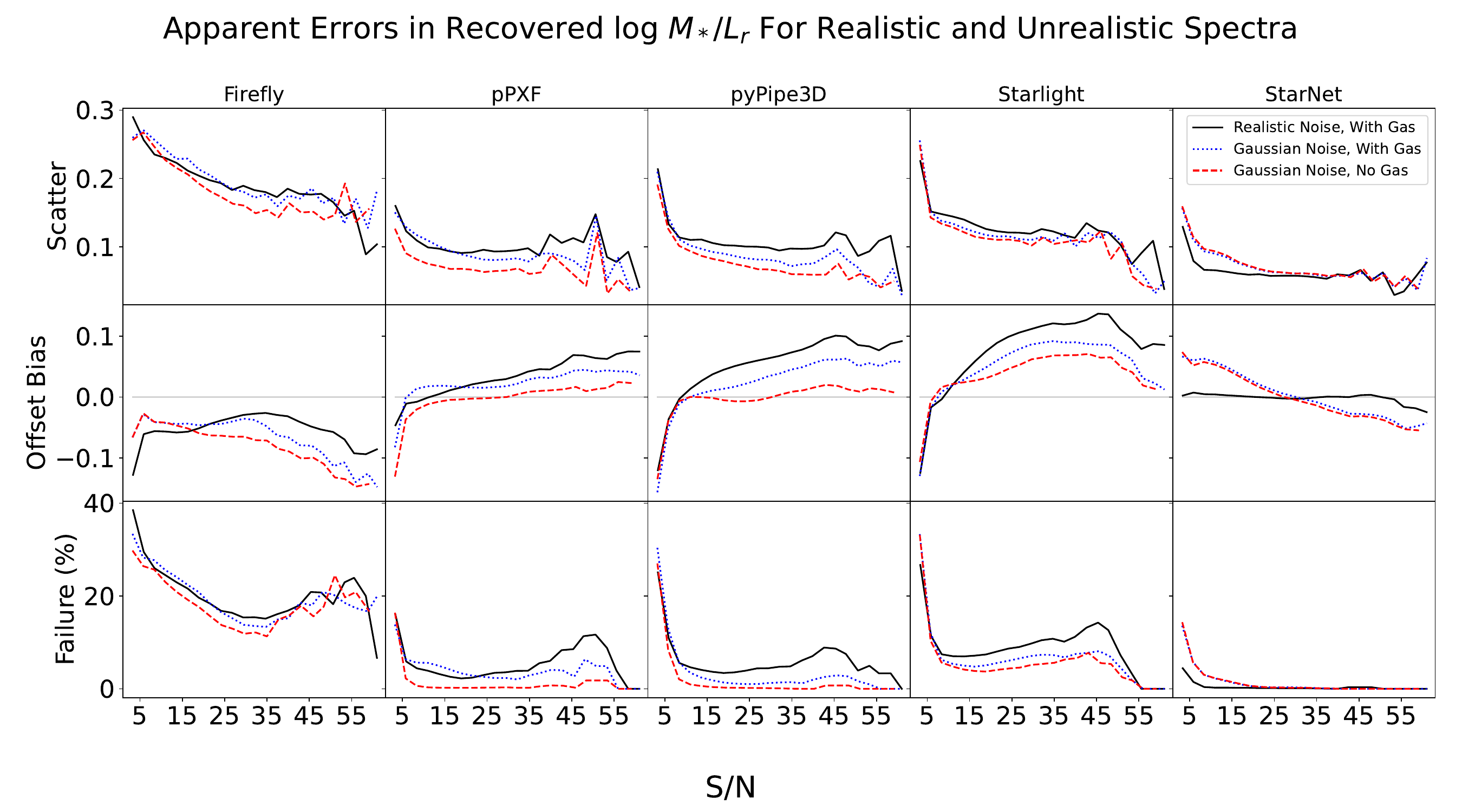}
    \caption{The effect of noise and gas emission lines on the recovered stellar mass-to-light ratio, namely the scatter in the errors, the mean offsets, and the failure rate as a function of mean S/N in each grid cell for the five tested codes.  Emission lines seem to have a bigger effect on the recovery of stellar $\MLr$ than the type of noise applied (the red dashed curves are more offset from the fiducial solid black curves than the blue dotted curves).}
    \label{errorsGas_lML}
\end{figure*}

\begin{figure*}
    \centering
    \includegraphics[width=\linewidth]{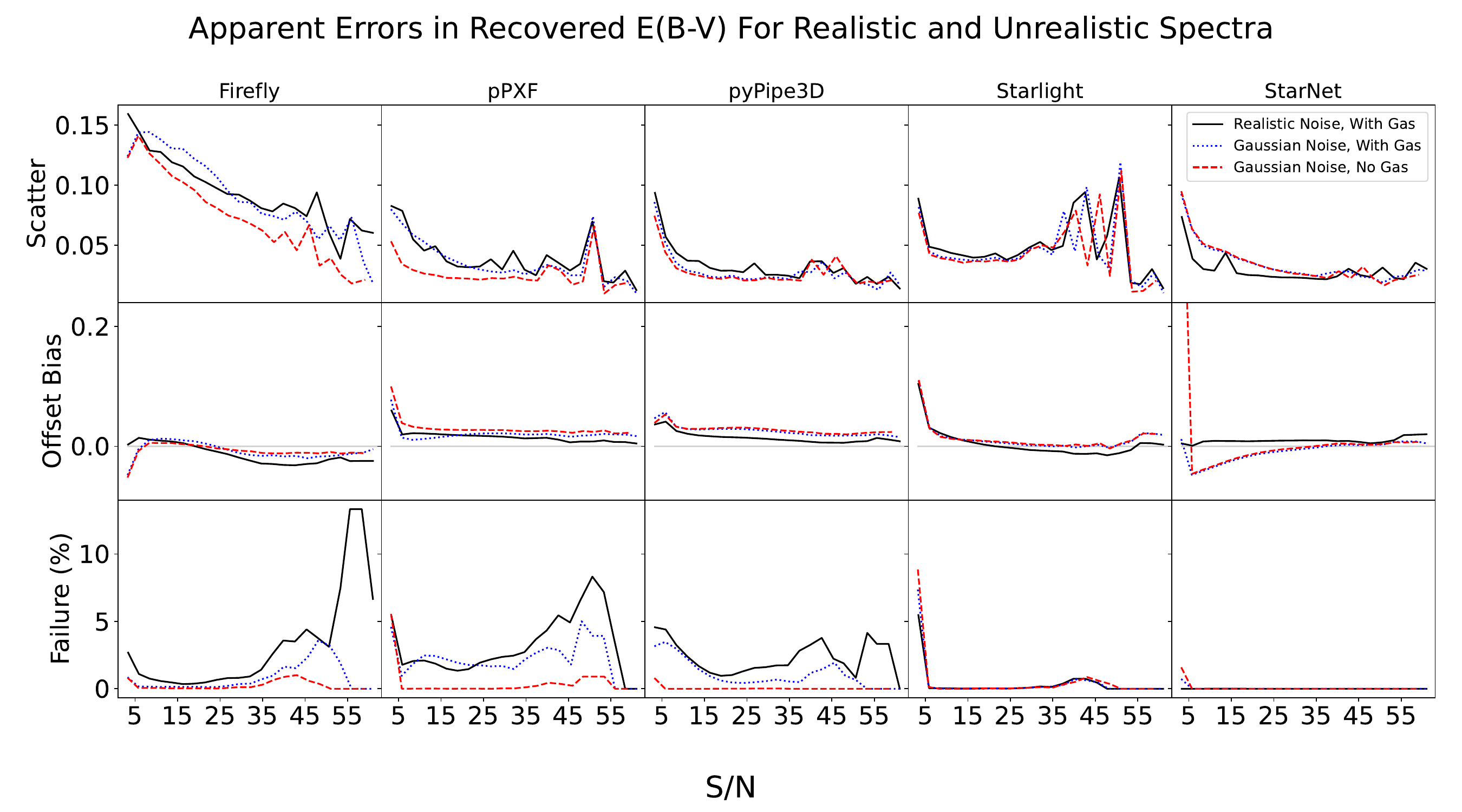}
    \caption{The effect of noise and gas emission lines on the recovered stellar E(B-V), namely the scatter in the errors, the mean offsets, and the failure rate as a function of mean S/N in each grid cell for the five tested codes.  The recovery E(B-V) seems not to depend much on the type of noise applied, or the presence of emission lines (the red dashed curves and blue dotted curves mostly stay close to the fiducial solid black curves).}
    \label{errorsGas_ebv}
\end{figure*}

\subsection{StarNet}
\label{starnet}

StarNet \citep{Fabbro2018} is a Convolutional Neural Network (CNN) that was designed to analyse stellar spectra to derive their effective temperature, surface gravity and metallicity. The architecture of the CNN consists of two convolutional layers, a ``max pooling" layer and three fully connected layers.  We refer to \cite{Fabbro2018} for details, but for readers who are less familiar with CNNs, we briefly describe what the layers do.

The first convolutional layer takes the spectrum as input and detects features by multiplying the spectrum with a set of filters (initialized with random shapes) of a user-defined length.  This process produces an activation spectrum where high values correspond to features in the original spectrum that are similar in shape to the filters.  For example, if one of the filters is a negative Gaussian, the activation spectrum will consist of strong signals where the original spectrum had absorption features.  The activation spectrum is passed onto a second convolutional layer which identifies features in the features.  An example of a feature in the features might include that certain of the original features occur in groups.  Including deeper convolutional layers like this is what makes CNNs a type of ``Deep Learning".  StarNet happens to be two (convolutional) layers deep.  

A max pooling layer selects only the strongest activations (\ie, the features of the features) within a window length of the second convolutional layer of the user's choosing, and passes them onto the three fully connected layers.  The purpose of this kind of selection and reduction of the dimensionality of the spectrum is to extract general features that are robust to small changes in the pixel values, such as would be caused by noise and even redshift.  
This is especially useful for recognizing a variety of spectral features in a dataset with a range of redshifts and noise profiles, such as our spectra, which imitate the range of redshifts and noise in the SDSS.

Fully connected layers (also called ``dense layers") are a set of linear functions called ``neurons" that map an input to an output.  The input to the first layer of neurons are the pooled features (of the features) detected by the convolutional layers.  The input for each neuron in subsequent layers are the outputs from every neuron from previous layers (hence the moniker ``fully connected").  The output values from the final layer are the desired stellar populations properties (mean log stellar age, mean stellar metallicity, E(B-V) and log $M_*/L_r$).  At their core the fully connected layers are a multi-parameter linear function that maps the detected features of the spectrum to the desired stellar population properties.

The numbers of filters, the filters lengths, and the number of neurons in the fully connected layers are called the ``hyperparameters", and these are set by the user.  We have used the default settings for StarNet except for the filter length, where we chose 20 (instead of 8) corresponding to about 14-34\AA, and a max pooling length 10 (instead of 4).  We chose to increase the filter length (and hence the max pooling length) because galaxy spectra have intrinsically wider features due to the higher velocity dispersions, compared to stellar spectra.  Furthermore, we have chosen to use both the galaxy spectra and the noise spectra as a 2-channel input to the StarNet to train the CNN to recognize when certain features are noisy and not be weighted heavily.

The shapes of the filters and the parameters of the fully connected linear neurons comprise  approximately 1.9 million free parameters (given our chosen hyperparameters) that the CNN initializes randomly.  The CNN is trained to find the appropriate values of these parameters that turn the input spectra into the four stellar population properties (stellar age, metallicity, E(B-V) and $\MLr$) by studying the training set spectra for which the true values of the stellar population properties are known.  Training is accomplished by adjusting the free parameters until the output consistently matches the true values to within a certain tolerance.  The loss function that measures success is the mean squared error, which is the quadrature sum of the difference between the predicted and true values of our 4 chosen stellar population properties (which were normalized to a range of 0 to 1).  The gradient of the loss function with respect to the free parameters is computed one layer at a time starting from the last layer and working backwards to the first in a process called ``backpropagation".

In contrast to the spectrum fitting codes, StarNet does not use SSP templates to derive stellar population properties. StarNet is in essence a complex multi-parameter function that computes 4 numbers from an input spectrum.  In principle, one could train StarNet to output many more than just four numbers, and we are investigating the abilities of StarNet to reproduce entire SFHs (Walters et al., in prep.).  Although StarNet does not require a prior in the form of a set of SSP templates, it does require a substantial training set, which we constructed from SSP templates and TNG SFHs.  Therefore, it is important to stress that StarNet is trained to recognize SFHs only from the TNG universe, and only through the linear combination of the E-MILES SSPs.  We settled on a training set of 200 000 spectra since our tests did not yield substantial improvements in parameter recovery for larger samples.

Since StarNet runs on a GPU, it required 9 minutes to train on 200000 spectra (done once), and thereafter took less than 0.0004s per spectrum to compute the four stellar population properties we tested.

\section{Results}
\label{results}

\subsection{Recovery of Four Test Properties of the Stellar Population}
\label{recovery}

Given the conditions described above, we ran the four spectrum fitting codes and one CNN on the mock spectra constructed in \secref{simulatedspectra}.  We test each code's ability to recover the mean of {the logarithm of} the mass-weighted stellar age,
{(i.e., $\langle \log t\rangle$)}, 
the mean mass-weighted stellar metallicity, the stellar mass-to-light ratio in the SDSS $r$-band and stellar colour excess E(B-V).  Our results are presented in \mfigs{lage}{ebv}.  Each of these figures shows the error in the predicted values (i.e., the predicted minus the true value) of the quantity in question, vs. the true value of the quantity (top panels) and the predicted values (bottom panels).  The black contours shows the distribution of points in the plot while the colour scale represents the mean S/N.  The scatter and mean offset biases (also called the bias) are indicated in the top panels, and are the same in the bottom panels.

A successful recovery of the test quantities means that the error in the predicted values is close to 0, which is marked by a dashed horizontal line in each plot.  One could debate what error threshold constitutes a successful recovery or a failure, and the answer likely depends on the science case for which these quantities are used.  For the sake of comparison between the codes, we defined a ``failure" to be an error of greater than 25\% of the full range of true values.  This limit is marked by the two dotted lines in each panel.  The points falling outside of these limits are ``failures" and the failure rates are indicated in all panels of \mfigs{lage}{ebv}. 

For all the quantities studied, StarNet recovered the stellar population properties with the smallest scatter ($< 0.08$ dex for all quantities), the lowest biases ($< 0.02$ dex for all quantities), the least systematic dependence of the bias on either the true or predicted values, the lowest failure rates ($<$ 1\% except for [Z/H] which was 2.7\%) and by far the fastest run time.  These results show that CNNs have great potential as a tool for determining stellar population properties from optical spectra.  The limitations of CNNs are well-known and will be discussed in \secref{discussion}.

Among the conventional spectrum fitting codes, pPXF has the smallest errors in recovering all our test quantities (scatter of $< 0.11$ dex, average bias of $< 0.08$) with the exception of E(B-V), which pPXF recovered with slightly worse scatter than pyPipe3D (0.038 vs 0.034 dex).  Given the superior speed of pPXF as well as its low errors, we conclude from these tests that pPXF is the best code overall (out of the conventional spectrum fitting codes) for recovering mean stellar population properties, within our experimental setup.  Firefly had the worst performance of all the codes, having the largest scatter and highest failure rates in the recovery of the test properties.  We discuss possible reasons for this in \secref{discussion}.
 
The bias depends on both the true and predicted values of the four properties for most of the codes.  For example, the tilt of the contours in the top panels \fig{metal} shows that all codes slightly overpredict the metallicities of the populations with the lowest true metallicities.  Given that we do not know the true values in real data, the lower panels show the measurement bias as a function of the measured (predicted) values.  \fig{lage} and \ref{metal} show that the biases in age and metallicity are strongly dependent on the predicted values for all the codes.  Even StarNet is not immune to these systematics, but the CNN had the weakest systematic dependence between the bias and the true and predicted values compared to the conventional codes.  For log $\MLr$ and E(B-V), the correlation between the bias and the true/predicted values is weaker, except perhaps for Firefly's measured log $\MLr$.

The black curves in \mfigs{errorsGas_lage}{errorsGas_ebv} show the scatter, offset and failure rate as a function of S/N for log age, [Z/H], log $\MLr$ and E(B-V) respectively.  (The blue and red curves will be discussed in \secref{gaussiannoise}.)  Most of the codes predict values that are on average biased (offset) from the true values by about 0.1 dex in [Z/H] (\fig{metal}), but they can approach 0.2 dex for high S/N spectra (\fig{errorsGas_metal}). The biases are 0.08 dex or smaller for stellar age and log $\MLr$ (\twofigs{lage}{lML}), but can reach as high as 0.15 dex for higher S/N spectra (\twofigs{errorsGas_lage}{errorsGas_lML}).

{Note that we have chosen to test mass-weighted quanitities since these are more physical (and are directly taken from the simulation), while the light-weighted quantities are more directly related to the data.  However we did perform the same tests for the light-weighted quantities and found very similar results for the metallicities, and only minimal improvements in predicting light-weighted age.}

\subsection{The Degeneracies Betweeen Age, Metallicity and Reddening}

\mfigs{lage}{ebv} are colour-coded by the mean S/N of the spectra in the pixels.  For the most part, the errors in the predicted stellar population quantities depend on S/N as we expect: regions farthest away from the zero lines (the highest errors) in \mfigs{lage}{ebv} also have the lowest S/N on average (red pixels in those plots).  However, there are some cases where the codes performed poorly on spectra even with high S/N.  For example, \fig{lML} shows that Firefly, pPXF and pyPipe3D all perform poorly for some high-S/N spectra with low true $\MLr$ (blue pixels in those panels have large errors).

\begin{figure*}
    \centering
    \includegraphics[width=\linewidth]{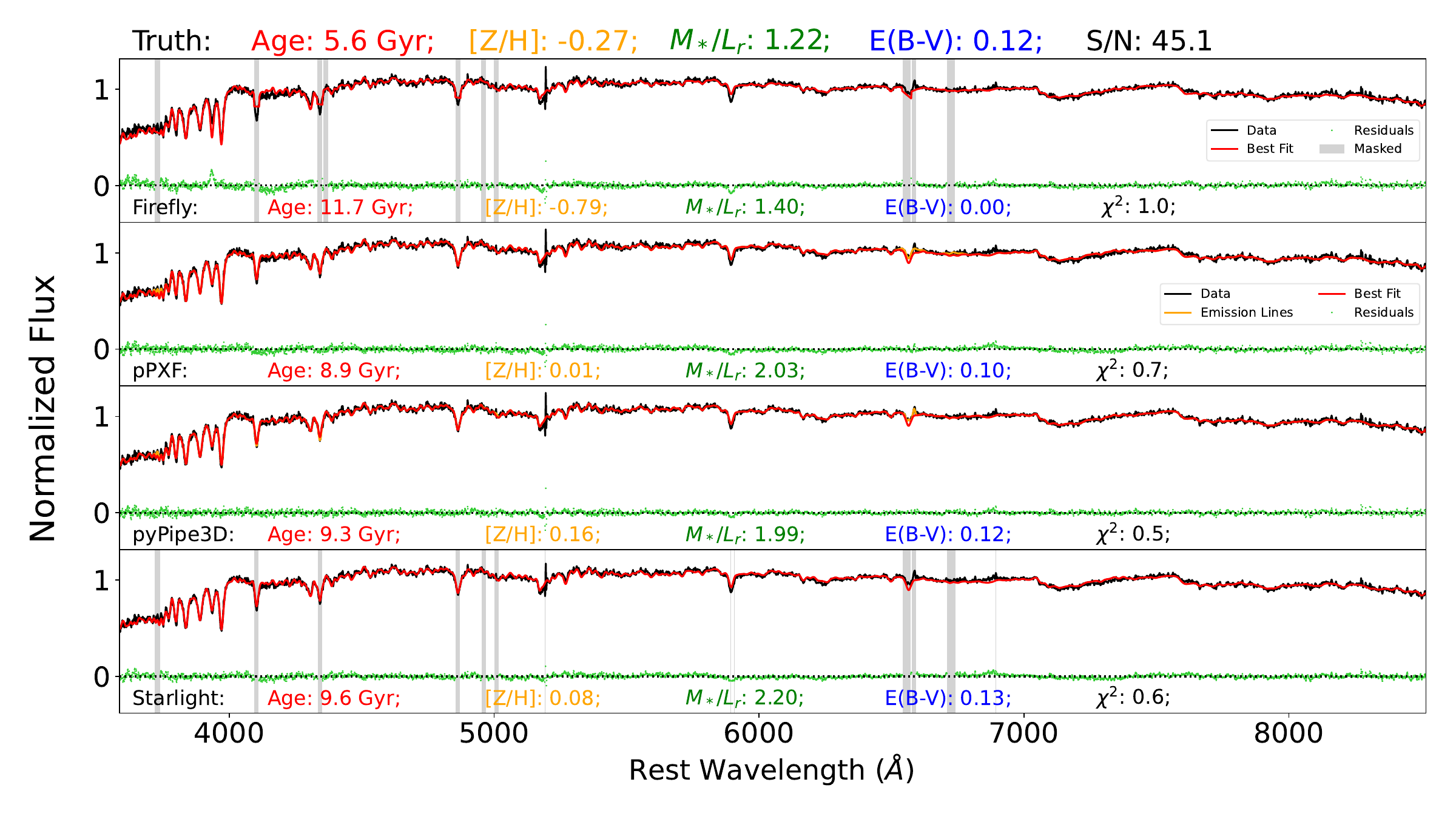}
    \caption{An example spectrum with high S/N for which the four non-DL codes produced good fits with acceptable $\chi^2$ (per degree of freedom), and yet had large errors in their measured parameters.  The four panels show the data and the fits for the four non-DL codes.  The true stellar population parameters are listed in the title, while the measured values for each code are listed in each panel.  Despite the acceptable $\chi^2$ for all the fits, the codes predicted ages and metallicities that were significantly wrong.}
    \label{examplefit}
\end{figure*}

Why do some of the parameters have large errors?  \fig{examplefit} shows an example spectrum that was fit by the four conventional codes.  The fitted parameters have large errors in age, metallicity and $\MLr$, despite the fact that all four fits to the SED have acceptably small $\chi^2$ errors.  The obvious culprit for these large errors are the known degeneracies in optical spectra, particularly between age, metallicity and dust reddening, three of the four quantities we are testing.

\begin{figure*}
    \centering
    \includegraphics[width=\linewidth]{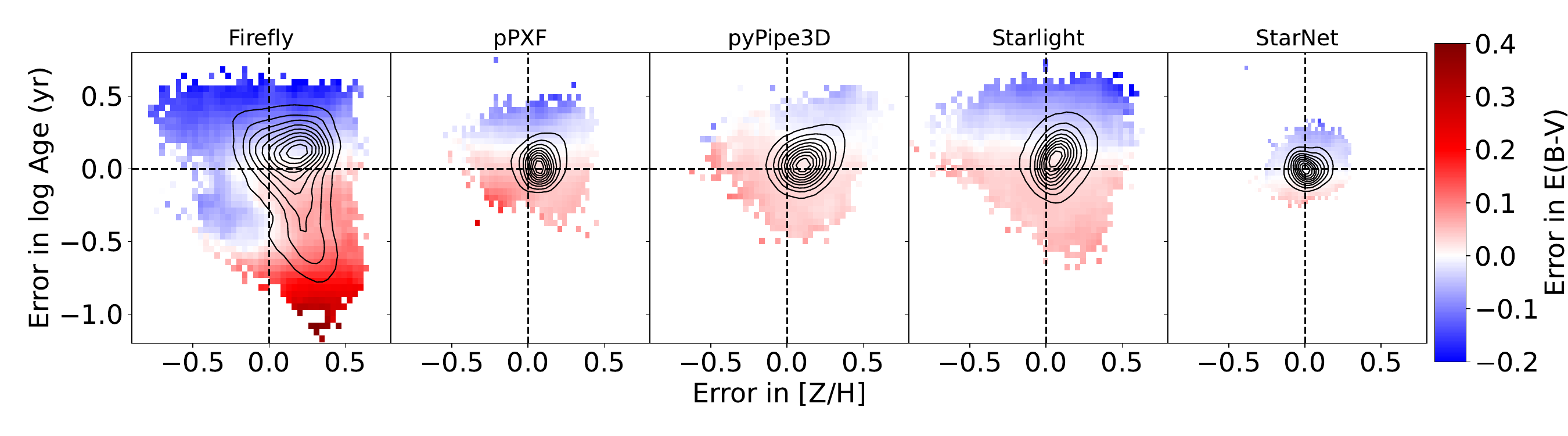}
    \caption{Plots exploring the age-metallicity-reddening degeneracy of each code.  The $y$-axis is the measured-minus-true log Age, and the $x$-axis is the measured-minus-true [Z/H].  The contours represent number density of points in the plot, while the colour scale shows the mean measured-minus-true E(B-V).  Firefly suffers the most from the age-metallicity-reddening degeneracy while the other codes suffer somewhat from a degeneracy between age and colour excess.}
    \label{degeneracies}
\end{figure*}

In order to examine how degenerate these quantities are for all tested codes, we show the errors in predicted stellar age versus the errors in predicted stellar metallicity in \fig{degeneracies}, colour-coded by the errors in the predicted E(B-V).  The contours show the distribution of points on the plot.  The classic age-metallicity degeneracy manifests if the contours show that the errors in age are anti-correlated with the errors in metallicity.  The contours for pPXF (once the [Z/H] offset is taken into account) and StarNet show that they are slightly affected by the classic age-metallicity degeneracy, although their contours are tight, reflecting the small scatter in their errors.  Firefly is a bit more strongly affected by the age-metallicity degeneracy, showing negative errors in age when the errors in metallicity are positive.  For pyPipe3D and {\sc starlight} the errors in age and metallicity are instead positively correlated (though the effects are small for both).  %*** no idea how to explain this ****

The colour code in \fig{degeneracies} shows how the errors in E(B-V) depend the errors in age and metallicity.  For the most part, errors in E(B-V) are small (see \fig{ebv}), but for all codes they depend on the errors in age and/or metallicity.  For all codes, errors in E(B-V) depend on the errors in age, and in the expected manner: if they overpredicted the mean stellar age, they also underpredicted the E(B-V).   All codes except Firefly were able to predict metallicity separately from E(B-V), likely recognizing the spectral indices.  However for Firefly, the errors in E(B-V) depend on both age and metallicity.  For Firefly, if the population age was underpredicted, then errors in E(B-V) depend positively on errors in metallicity.  

{We note that Firefly is the only code that leaves the reddening law completely free, while the other codes assume the same law \citep{Calzetti2000} that was applied to the mocks in the first place, which may have a significant impact on the recovery of E(B-V).} 

\subsection{The Importance of Realistic Mock Spectra}
\label{gaussiannoise}

What difference does it make to use realistic noise and/or emission lines in these tests?  
The effort to produce a suite of mock spectra that closely mimic the non-Gaussian noise and emission lines of the SDSS was non-trivial.  Was it worth it?  To answer this, we created two more sets of 50000 mock spectra that had identical SFHs to our fiducial set, but one set used Gaussian noise instead of SDSS residuals, and the other used Gaussian noise AND excluded the emission lines.  We ran the five codes on these additional mock spectra and compared their results.  \mfigs{errorsGas_lage}{errorsGas_ebv} show the scatter, offset and failure rate as a function of S/N for log age, [Z/H], log $\MLr$ and E(B-V) respectively.  The results for the fiducial set of mock spectra are shown as solid black lines (labeled ``Realistic Noise, With Gas"), the results for spectra with Gaussian noise are the dotted blue lines, and the result for spectra  with Gaussian errors and no emission lines are the dashed red lines.

By comparing the red dashed and blue dotted curves with the solid black curves in \mfigs{errorsGas_lage}{errorsGas_ebv}, it is clear that neglecting realism in mock spectra can lead to significant underestimation in the errors of the recovered stellar population properties.  The largest underestimates are in the recovery of the stellar metallicity [Z/H] (\fig{errorsGas_metal}).  If one were to use Gaussian noise instead of realistic noise, one would conclude that the scatter in the recovered [Z/H] values is only half as bad than is realistic.  The red and blue curves in \fig{errorsGas_metal} show that the bias and failure rate appear to be nearly 0 for spectra with high S/N when in reality they should be significantly higher (the solid black curves).  Therefore, by ignoring realistic noise, parameter recovery tests may significantly underestimate errors in [Z/H].

If one uses Gaussian noise, the addition or omission of gas emission lines makes little difference in the recovery of [Z/H] (the red and blue curves are similar in \fig{errorsGas_metal}).  The unimportance of emission lines to [Z/H] is not surprising because the main markers for stellar metallicity (e.g., the Mg and CaII absorption lines) are not usually contaminated by emission.  Where gas emission makes more of a difference is in the recovery of stellar age, and the related $\MLr$ (\twofigs{errorsGas_lage}{errorsGas_lML}).  The difference is most pronounced for pPXF, pyPipe3D and {\sc starlight}.  The scatter in $\lMLr$ can be underestimated by as much as a factor two, and by as much as 50\% for stellar age.  The offsets and failure rates for both age and $\lMLr$ are misleadingly close to zero if no emission lines (and Gaussian noise) are added (\fig{errorsGas_lML}).  This makes sense since some of the main markers of stellar age, namely the Balmer absorption lines, are often contaminated by gas emission.  

\fig{errorsGas_lML} shows that the recovery of stellar E(B-V) is affected very little by the presence or absence of emission lines, or by the use of realistic vs Gaussian noise (since the red and blue curves mostly follow the solid black curves).  This is understandable since reddening affects the shape of the continuum, which is relatively insensitive to the type of noise or the presence of emission lines. 

For the StarNet CNN trained on realistic spectra, running the model on spectra with Gaussian noise and emission lines made very little difference in the scatter and failure rates of all four stellar population parameters.  However, the mean offsets of the recovered parameters (except for E(B-V)), {\it worsen} with the simpler spectra.  This is perhaps a sign of overfitting in the training (more on this in \secref{discussion}), and further study is needed before StarNet can be used on real spectra.

\section{Discussion}
\label{discussion}

In our tests of stellar population parameter recovery, we have created mock spectra using realistic noise and emission lines, which turns out to be important for estimating errors and biases in the recovered parameters.  We also found that the StarNet CNN significantly outperforms the conventional codes in both parameter recovery and computation time.  Out of the conventional spectrum fitting codes, pPXF had the best performance overall in both parameter recovery and computation time.  

CNNs are known to outperform traditional algorithms in many applications (\citealp{Huertas-Company2018,Lovell2019,Brescia2021,Teimoorinia2022}).  Our results add to the growing literature demonstrating the significant potential of CNNs to aid astronomy in an age of ever-increasing data and computational demands.  The limitations of machine learning are well-known, including the need for a training set, and the lack of an error estimate on the derived parameters {(although see \citealp{Bialek2020} on estimating uncertainties for CNNs using an ensemble of models)}.  Furthermore, it is unknown how generalizable the trained model is to data that are dissimilar to the training set.  In our case, StarNet was trained to recognize SFHs drawn from TNG100 and it is unclear how realistic these are, and how well StarNet can derive stellar population properties from spectra in the real universe.  We have already seen signs of overfitting in \secref{gaussiannoise}, namely that StarNet trained on realistic spectra show {\it worse} offset bias when run on simpler spectra.  We remind the reader that, although the training set consists of 200 000 spectra, they are created from only 20962 unique SFHs. 
However we include StarNet in this study to demonstrate its potential to outperform conventional tools in every respect, and to motivate the study of the limitations of CNNs in more detail (Walters et al, in prep.).  One of our future goals is to produce a model-independent training set, the SFHs of which produce spectra with the same range of continuum properties as observed (i.e., not just in noise and emission lines).  (For a model-dependent ``forward-modelling" approach to this problem, see \citealp{Sarmiento2023}.)

In our tests, Firefly had the largest scatter, bias offsets and failure rates in the recovery of our chosen stellar population parameters (but was the 2nd-fastest of the conventional codes).  As explained in \secref{firefly} we had to modify the code to use the E-MILES templates, since the mock spectra were created from E-MILES and since we used the E-MILES templates in all the other codes.  To check that this modification was the not the source of Firefly's performance, we created another set of mock spectra using the the MaStar templates, and re-ran Firefly using the MaStar templates to fit these new spectra.  We found similar values in the scatter, bias offset and failure rates for all four quantities tested.  {A similar recovery test by \cite{Nanni2023} with mock MaNGA spectra (but using Gaussian noise) found similar scatter and bias offsets.}
We suspect that the culprit behind Firefly's relatively high errors is the use of SFHs that are a combination of SSPs with {\it equal} weights (in light).  The resulting best-fit SFH is constructed by weighting these unrealistic SFHs by their likelihood, producing a final more realistic-looking SFH (i.e., one that has non-uniform weights on the SSPs).  However, using only combinations of uniform SFHs (for runtime considerations) is perhaps too restrictive and likely misses non-uniform SFHs that have higher likelihood.  %We defer a deeper exploration of Firefly's fitting mechanisms to a future study.

{\sc starlight} ranks third in terms of parameter recovery out of the four conventional spectrum fitting codes.  It had relatively high scatter and failure rates in the recovery of all parameters except E(B-V).  \cite{CidFernandes2018} suggested that it can take the Markov chains in {\sc starlight} a longer time to reach the extreme corners of the $\chi^2$ parameter space, such as highly reddened populations, and therefore for such galaxies, the ``slow" mode of {\sc starlight} might be more appropriate.  Indeed we have found that it is the youngest populations with low metallicity and high E(B-V) that have the largest errors.  
{Note that {\sc starlight} in the default mode was already the slowest of the codes we tested.  However we note that there is a non-public version of {\sc starlight} that is significantly faster (R. Cid Fernandes, private communication).}  

pyPipe3D ranks second in our tests in terms of parameter recovery.  Like {\sc starlight}, pyPipe3d also uses an MC method to solve the SFH part of the problem, but has a similar accuracy to pPXF in recovering our 4 stellar population parameters.  Therefore, MC methods do have the potential to accurately recover stellar population properties, with the major added benefit of estimating the parameter errors using the exploration of the parameter space.  However the cost is speed, as pyPipe3D was the {second-slowest} of the tested codes, being an order of magnitude slower than pPXF.  (Note that Pipe3D was originally written in faster languages, but since pPXF was also written in Python, the speed comparison is fair.)  The fact that pyPipe3D had slightly worse parameter recovery compared to pPXF may be due to the fitting of the SFH separately from the kinematics, the emission lines and the reddening, compared to simultaneous fit of all parameters in pPXF.  

pPXF had the best performance out of the four non-DL codes we tested in both recovery of our four chosen parameters and in speed.  Its speed is due to the exploitation of the quadratic nature of the $\chi^2$ merit function and the use of well-known efficient least-squares solutions to such problems.  The ability to regularize the SFHs is a major additional benefit to pPXF.  {In fact, we also tested pPXF without regularization enabled and found that pPXF's parameter recovery was comparable to that of {\sc starlight}'s.  Parameter recovery generally improved for higher values of the regularization (``\texttt{regul}") parameter, but improvements were minimal beyond \texttt{regul=100}.  Therefore regularization results in a significant improvement to the recovery of stellar population parameters, and appears to be the reason that pPXF had the best parameter recovery (aside from the StarNet CNN).}

Yet pPXF and even StarNet, along with the other codes, show systemic dependence between the bias errors and the true and predicted values of the stellar age, metallicity and $\MLr$.  Metallicity is especially difficult for all codes to reproduce, and our study underscores the caution needed when inferring stellar population properties via full spectrum fitting of optical spectra.

Although our suite of mock spectra include realistic noise and emission lines, our tests of parameter recovery are still only lower limits on the parameter errors, offset biases, and failure rates of the tested codes for a number of reasons.  Firstly, we have used (a subset of) the same SSP spectra that were used to create the mock spectra.  Real spectra likely consist of more complex populations, perhaps with different IMFs and alpha-enrichments.  Secondly, we have not taken any AGN contribution to the continuum into account, even though broad-line and narrow-line AGN emission has been included when SDSS emission lines and residuals were added to the mock spectra.  Thirdly, we have also assumed TNG SFHs which may or may not be a realistic representation of the universe.  Therefore the statistical measures of the success the parameter recovery may not represent the rate of successful parameter measurement in the real universe.

\section{Summary and Conclusions}

In summary, we have tested the ability of several spectrum fitting codes (Firefly, pPXF, pyPipe3D and {\sc starlight}), and one DL code (StarNet) to recover four stellar population parameters: the stellar age, the stellar metallicity, the stellar mass-to-light ratio, and the colour excess due to reddening.  For this purpose, we created a realistic set of mock spectra constructed from SFHs drawn from the IllustrisTNG simulation, the E-MILES SSP templates and real noise profiles and emission lines from observed spectra in the SDSS.  Our main findings are as follows:
\begin{enumerate}
 \item The StarNet CNN vastly outperforms the conventional codes in both parameter recovery (scatter in the errors of $< 0.08$ dex, biases $< 0.02$ dex for all quantities) and computation time (see \tab{comparisontable}).
 \item Of the conventional (non-DL) codes, pPXF had the best parameter recovery overall (errors of $< 0.11$ dex, biases $< 0.08$ dex), and also the fastest computation time.  pyPipe3D was a close second in terms of parameter recovery, but its computation time was almost an order of magnitude slower.
 \item Firefly suffers the most from the age-metallicity-reddening degeneracy while the other codes suffer somewhat from a degeneracy between age and colour excess.
 \item We have found that using realistic noise in mock spectra is most crucial for an estimate of errors in the stellar metallicity, where unrealistic Gaussian noise can lead to an underestimate of the errors by as much as a factor of two.  The addition of emission lines was important for the error estimation in stellar age and $\MLr$, where neglecting to add emission lines could lead to an underestimate of the errors in $\MLr$ by a factor of two, and in stellar age by 50\%.  Furthermore, systematic biases in age, metallicity and $\MLr$ can be missed entirely if using unrealistic noise or neglecting nebular emission in mock spectra.
 \item The recovery of E(B-V) did not depend significantly on the type of noise or the presence or absence of emission lines.
\end{enumerate}

Our results demonstrate the potential of CNNs as a tool in stellar population studies, but further study with larger, more diverse and realistic training sets, is needed to determine the generalizability of such tools.  Although we found pPXF to be the best of the conventional codes, all codes exhibit systematic biases, especially in the recovery of stellar metallicity, but also in stellar age and $\MLr$.  Furthermore, our tests have used several assumptions, including simulation SFHs and a particular set of SSPs.  Therefore our results represent a lower limit to the errors, offset biases and failure rates of the codes we tested.

\section*{Data Availability}

Our entire suite of mock spectra is available to download from \url{https://tiny.cc/Woo2024}.

\section*{Acknowledgements}
{We thank Roberto Cid Fernandes and Michele Cappellari for helpful comments that improved this paper.}
We acknowledge the helpful and stimulating discussions with David Koo, Joel Primack, David Elbaz, Marc Huertas-Company, Jing Wang and Sandro Tacchella.

We thank Spencer Bialek for sending us a copy of StarNet and for helpful comments, {as well as Sebastien Fabbro for helpful comments}.

This research made use of Astropy,\footnote{http://www.astropy.org} a community-developed core Python package for Astronomy \citep{Astropy2013,Astropy2018}. 
This research also made use of the computation resources provided by the Digital Research Alliance of Canada (formerly Compute Canada) (https://alliancecan.ca).  

Funding for the Sloan Digital Sky Survey IV has been provided by the Alfred P. Sloan Foundation, the U.S. Department of Energy Office of Science, and the Participating Institutions. SDSS acknowledges support and resources from the Center for High-Performance Computing at the University of Utah. The SDSS web site is www.sdss.org.  SDSS is managed by the Astrophysical Research Consortium for the Participating Institutions of the SDSS Collaboration including the Brazilian Participation Group, the Carnegie Institution for Science, Carnegie Mellon University, the Chilean Participation Group, the French Participation Group, Harvard-Smithsonian Center for Astrophysics, Instituto de Astrofísica de Canarias, The Johns Hopkins University, Kavli Institute for the Physics and Mathematics of the Universe (IPMU) / University of Tokyo, the Korean Participation Group, Lawrence Berkeley National Laboratory, Leibniz Institut für Astrophysik Potsdam (AIP), Max-Planck-Institut für Astronomie (MPIA Heidelberg), Max-Planck-Institut für Astrophysik (MPA Garching), Max-Planck-Institut für Extraterrestrische Physik (MPE), National Astronomical Observatories of China, New Mexico State University, New York University, University of Notre Dame, Observatório Nacional / MCTI, The Ohio State University, Pennsylvania State University, Shanghai Astronomical Observatory, United Kingdom Participation Group, Universidad Nacional Autónoma de México, University of Arizona, University of Colorado Boulder, University of Oxford, University of Portsmouth, University of Utah, University of Virginia, University of Washington, University of Wisconsin, Vanderbilt University, and Yale University.

\bibliographystyle{mnras}
\bibliography{references}
 
\label{lastpage}

\end{document}